\documentclass[preprint,12pt]{elsarticle}

\usepackage{graphicx}

\usepackage{setspace}

\usepackage[utf8x]{inputenc}
\usepackage[T1]{fontenc}
\usepackage[english]{babel} 
\usepackage{subcaption} 
\usepackage{textcomp}
\usepackage{amssymb} 
\usepackage{amsmath} 

\usepackage[onelanguage, ruled, lined, vlined]{algorithm2e}
\journal{ }

\begin{document}
\begin{frontmatter}
\title{Efficient spatial designs for  targeted regions or level set detection}

\author[inst1]{Sylvain Coly}
\author[inst1]{Pierre Druilhet}
\author[inst1]{Nourddine Azzaoui}

\affiliation[inst1]{organization={Laboratoire de Math\'ematiques Blaise Pascal},
            addressline={UMR 6620 CNRS},
            city={place Vasarely},
            postcode={63 178},
            state={Aubi\'ere Cedex},
            country={France}}

\date{ }
\begin{abstract}
Acquiring information on spatial phenomena can be costly and time-consuming.   In this context, to obtain reliable global knowledge, the choice of measurement location is a crucial issue.  Space-filling designs are often used to control variability uniformly across the whole space. However, in a monitoring context, it is more relevant to focus on crucial regions, especially when dealing with sensitive areas such as the environment, climate or public health. It is therefore important to choose a relevant optimality criterion to build models adapted to the purpose of the experiment.  In this article, we propose two new optimality criteria: the first aims to focus on areas where the response exceeds a given threshold, while the second is suitable for estimating sets of levels.   We introduce several algorithms for constructing optimal designs. We also focus on cost-effective algorithms that produce non-optimal but efficient designs.   For both sequential and non-sequential contexts, we compare our designs with existing ones through extensive simulation studies.
\end{abstract}

\begin{keyword}
 Optimal designs\sep  computer experiment\sep  kriging \sep  Gaussian processes.

\end{keyword}

\end{frontmatter}

\section{Introduction}

In many situations, collecting spatial data can be costly or time-consuming. This is the case in epidemic propagation studies, in the spatial control of pollutants, in climatic phenomena, and so on. When the number of observations is limited, it is crucial to optimize the positions of sampling locations, taking into account prior information, generally obtained from previous or indirect observations or diffusion models. In the case of sequential monitoring, the next sampling locations can be selected on the basis of up-to-date information. For single-stage sampling, we can only rely on prior information to select sampling positions.

Space filling designs are the most popular non-sequential designs when no prior information is available. They aim to ensure uniformly accurate estimation of the phenomenon over the whole space,  see, e.g. \cite{Niederreiter1987,Johnson1990}.
For example, distance-based approaches such as  \emph{Minimax}, \emph{Maximin} or other   discrepancy criteria have been proposed for the construction of   optimal space filling  designs.  To reduce computation costs,   low discrepancy sequences such as Halton, Hammersley, Sobol and Faure sequences have provided a first attempt to build easy to compute designs. For high dimensional problems, Latin Hypercube Designs and orthogonal arrays have been introduced to ensure equilibrium on axes \cite{Stein1987,Owen1992}.

In the context of Gaussian fields or Kriging methods,   entropy \cite{Shewry1987,Currin1991,buesoaa1998, anguloba2013} and   Integrated Mean Square Error (IMSE) \cite{Sacks1989} are the    two  main criteria  proposed for optimal sensor deployment.  Numerous alternatives  have been proposed in recent  decades: Conditional Minimizer Entropy (CME) \cite{Villemonteix2012},   Generalized Relative
Complexity \cite{alonsoba2016}, Expected Improvement (EI) \cite{Jones1998} and   Expected Improvement Gain (EIG) \cite{Ryan2003}. All these methods aim to determine the design carrying the greatest expected amount of information \cite{MateuMuller2012}. Recently, many variants of   Expected Improvement (EI) \cite{Bichon2008,Ranjan2008}, such as the quasi-Expected Improvement (q-EI) \cite{Chevalier2012} have been introduced and widely  used in industrial fields.

The approaches cited above aim to control the global variance without considering the expected values of the response.
The aim of the paper is to propose optimal designs based on criteria targeting an area of interest. The approaches cited above aim to control the global variance without considering the expected values of the response.
The aim of the paper is to propose optimal designs based on criteria targeting an area of interest. We mainly consider two cases: the first aims to target regions where the response values is significantly high. The second, more suited to sequential designs, aims to estimate a given set of levels.

   In section \ref{section.criteria}, we motivate and define the optimality criteria used to target regions of interest.
  In section \ref{section.algorithm}, we propose some algorithms for obtaining efficient designs.
  In section \ref{section.seqdesignexample}, we carry out simulation studies for sequential designs when the target area is a level set. We propose several performance indicators to compare our designs with those obtained in \cite{Picheny2010}. We then discuss the relative merits of each method.
  In section \ref{section.nonseqstudy}, we computationally evaluate the effectiveness of our methods for single-stage designs on a few examples.

\section{Optimality criteria  focusing on areas of interest}
\label{section.criteria}

We consider a grid $E$ of size $N\times N$, where $N$ is an integer. We denote by $y(x)\geq 0$ the variable of interest at  $x\in E$, and by $y$ the $N^2-$vector with entries $y(x)$. We consider a grid $E$ of size $N\times N$, where $N$ is an integer. We denote by $y(x)$ the true response at point   $x$ in $E$, and by $y$ the $N^2-$vector with entries $y(x)$.We assume that knowledge of $y$ can be modeled by a Gaussian field:

\begin{equation}
	\label{eq.metamodel}
	y\sim\mathcal{N}(\mu,\Sigma),
\end{equation}
which corresponds to prior knowledge from a Bayesian perspective, or, in the field of computer experiments,  to a meta-model.  For $x\in E$, we denote $\mu(x)=\mathbb E(y(x))$ and $ \sigma^2_x=\Sigma_{x,x}=\mathrm{Var} (y(x))$.

Let  $d=\{x_i\}_{1\le i\le n}$ be  a  $n$-point design and   $y_d=(y(x_i))_{1\le i\le n}$  be  the $n$-vector $y$ restricted to $d$. Knowing  the response $y_d$ on the design points, the updated knowledge on the field is given by the linear updating  formula:

\begin{equation}
	\label{updatedmean}
	\mu_{\bar d|y_d}= \mathbb E(y_{\bar d}|y_d)=\mu_{\bar d}+\mathrm{Cov}\left(y_{\bar{d}},y_d\right)\;\mathrm{Var}^{-1}\left(y_d\right)(y_d-\mu_d),
\end{equation}

and for the variance:
\begin{equation}
	\label{updatedvar}
	\mathrm{Var}\left(y_{\bar{d}}|y_d\right)=\mathrm{Var}\left(y_{\bar{d}}\right)-\mathrm{Cov}\left(y_{\bar{d}},y_d\right)\;\mathrm{Var}^{-1}\left(y_d\right)\;\mathrm{Cov}\left(y_d,y_{\bar{d}}\right),
\end{equation}
where $\bar d=E\setminus d$ is the complement of $d$, $y_{\bar{d}}$ the  vector of responses  restricted to $\bar d$ and  $\mu_d=\mathbb{E}(y_d)$.
Since  $\mathrm{Var}\left(y_{\bar{d}}|y_d\right)$ does not depend on the actual value of $y_d$  but only on the position of the observation points, we write $\mathrm{Var}\left(y_{\bar{d}}|d\right)$ instead of $\mathrm{Var}\left(y_{\bar{d}}|y_d\right)$. Similarly we write  $\mathrm{Var}\left(y(x)|d\right)$ instead of  $\mathrm{Var}\left(y(x)|y_d\right)$.

   In most applications, measurement errors are negligible compared to spatial variability. Therefore, we assume that  $\mathrm{Var}\left(y(x_i)|d\right)=0$ for  $x_i$ belonging to $d$ and  we will confound $y(x_i)$ with its measurement.

 The aim of the paper is to construct a design $d$ that provides accurate knowledge on the area of interest. In order to define optimality criteria,  we first  define for any $x$ in $E$   the weighted variance as in \cite{Picheny2010} :
\begin{equation}
\label{eq.cxd}
 c(x;d)=w(x)\times \mathrm{Var}\left(y(x)|d\right),
\end{equation}
where $w(x)$ is a  weight function that depends on the area of interest.  Two types of global criteria can be derived from this weighted variance function: a max-criterion

\begin{equation}
\label{eq.Maxcrit}
 \operatorname{MC}(d)=\max_{x\in E}  c(x;d),
\end{equation}
 and an integrated criterion
\begin{equation}
\label{eq.integratedcrit}
 \operatorname{IC}(d)= \sum_{x\in E} c(x;d).
\end{equation}

Then, we seek the design $d^*$ that minimizes either $\operatorname{\operatorname{MC}}(c)$ or $\operatorname{\operatorname{IC}}(d)$. In the next sections, we propose several weight functions, depending on the goal of the experiment and the target zone.  Note that a design $d^*$ that minimizes $\operatorname{MC}(d)$ also minimizes $\max_{x\in E}  h(c(x;d))$ for any   increasing transformation $h$. Such monotonic invariance property does not hold for $\operatorname{IC}(d)$.

 In the next sections, we propose two different weight functions, depending on the goal of the experiment and the target area.

\subsection{Level set detection}
\label{subection.crit3}

Here, we aim to estimate the level set $\mathcal L=\{x\in E\;;\;y(x)=T\}$ associated to a given threshold $T>0$. We propose the  weight function   defined by
\begin{equation}
\label{eq.weightpxls}
w^{\mathrm{ls}}_T(x)=2\left|\frac{1}{2}-F\left(\frac{\mu_x-T}{\sigma_x}\right)\right|.
\end{equation}
where $F$ is  the cumulative distribution function of the standard normal distribution.  There are two possible interpretations of the weight function  $w^{\mathrm{ls}}_T(x)$.
\\
The first one is frequentist:   consider  $y(x)$ as the unknown fixed quantity to be evaluated and $\mu(x)$ as the realization of a  normally distributed random variable with mean $y(x)$ and variance $\mathrm{Var}(y(x))$. In that case,   $w^{\mathrm{ls}}_T(x)$ corresponds to the p-value of the two-tailed  test :  $\mathcal H_0:``y(x)=T"$ vs $\mathcal H_1:``y(x)\neq T"$. The weight  $w^{\mathrm{ls}}_T(x)$ is close to $1$ when the hypothesis  "$x$ belongs to the level set"  is rejected and is close to $0$ otherwise.
\\
 The second interpretation is Bayesian:   put a flat prior on $y(x)$ that reflects the ignorance on $y(x)$. Consider the HPD-credible set $C_\beta(x)$ of  $y(x)$ with credible level $\beta$, then  $w^{\mathrm{ls}}_T(x)$ is the maximal credible level $\beta$ such that $T$ does not belongs to $C_\beta(x)$. This corresponds to the construction of Bayesian two-sided hypothesis testing based on confidence intervals (see \cite {Lindley1965, Thulin2014})

We denote  by  $\operatorname{MC}^{\mathrm{ls}} (d)$ and $\operatorname{IC}^{\mathrm{ls}} (d)$  the max and integrated  criteria related to $w^{\mathrm{ls}}_T(x)$.

 ~

In \cite{Picheny2010}, another weight function has been proposed to estimate level sets. It is denoted by  $w_{\sigma^2_\epsilon}(x)$  in this paper and  defined by

\begin{equation}
\label{eq.weightpicheny}
w_{\sigma^2_\epsilon}(x)=\frac{1}{\left( {2\pi}(\sigma^2_\epsilon+\sigma^2_x)\right)^{1/2}}\exp\{-\left(\mu_x-T\right)^2/\left(2\left(\sigma^2_\epsilon+\sigma^2_x\right)\right)\},
\end{equation}
where $\sigma^2_\epsilon$ is a  smoothing parameter that needs to be calibrated.

 We denote  by  $\operatorname{MC}^{ {W}}(d)$ and $\operatorname{IC}^{ {W}}(d)$  the max and integrated  criteria  derived from  $w_{\sigma^2_\epsilon}(x)$. The authors in \cite{Picheny2010} consider only $\operatorname{IC}^W$. We will show on examples that the performance of designs based on $\operatorname{MC}^{ {W}}(d)$ are less efficient than that based on $\operatorname{IC}^{ {W}}(d)$.

\begin{figure}
	\begin{center}
		\begin{subfigure}[b]{0.48\textwidth}
			\begin{center}
	\includegraphics[width=0.95\textwidth]{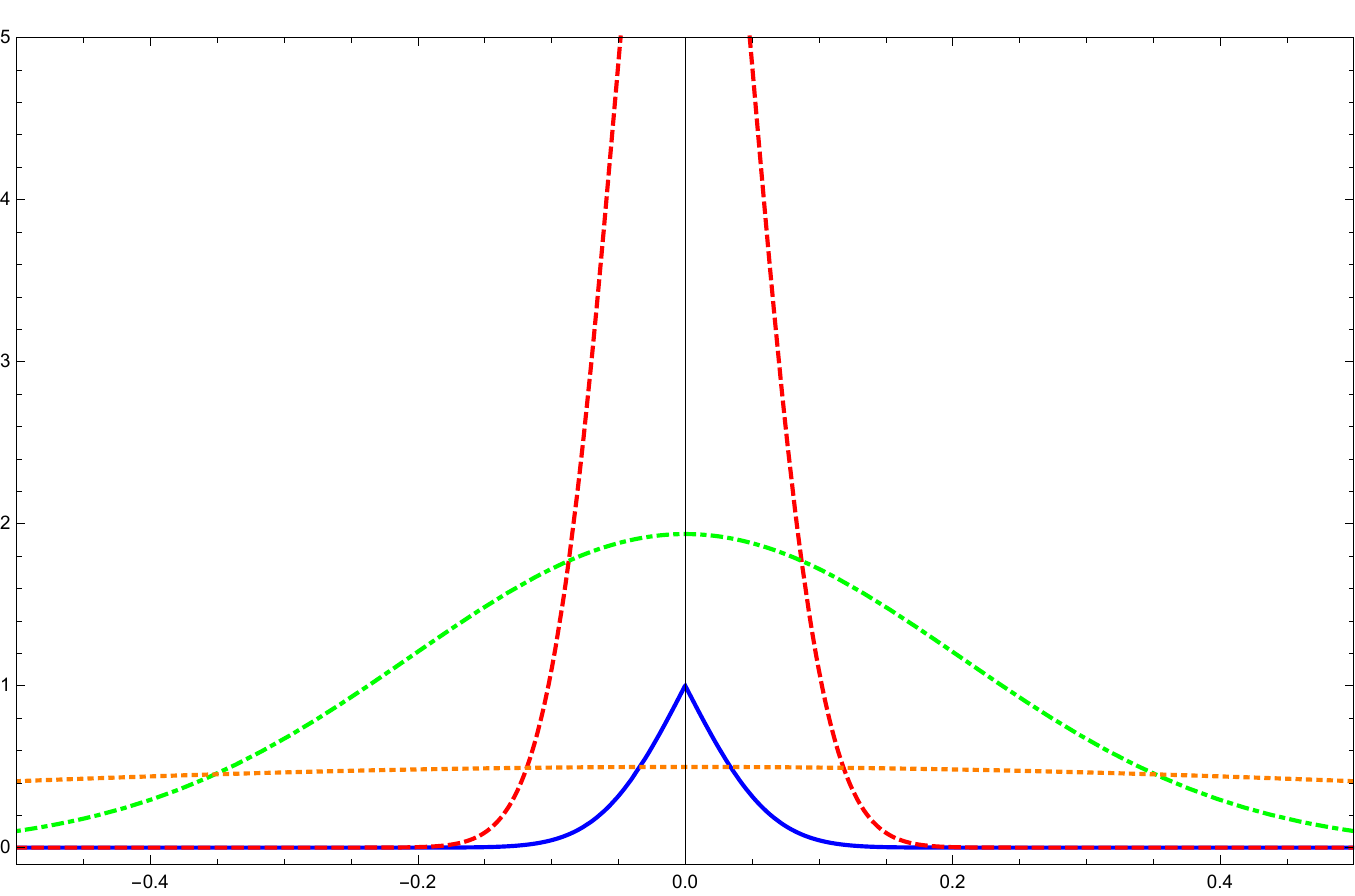}
	\captionsetup{justification=centering}
	\caption{weight functions w.r.t. $\mu_x-T$\\ for $\sigma_x=0.05$  }
 \label{fig.weightsfunctionsmean}
			\end{center}
		\end{subfigure}
		\begin{subfigure}[b]{0.48\textwidth}
			\begin{center}
	\includegraphics[width=0.95\textwidth]{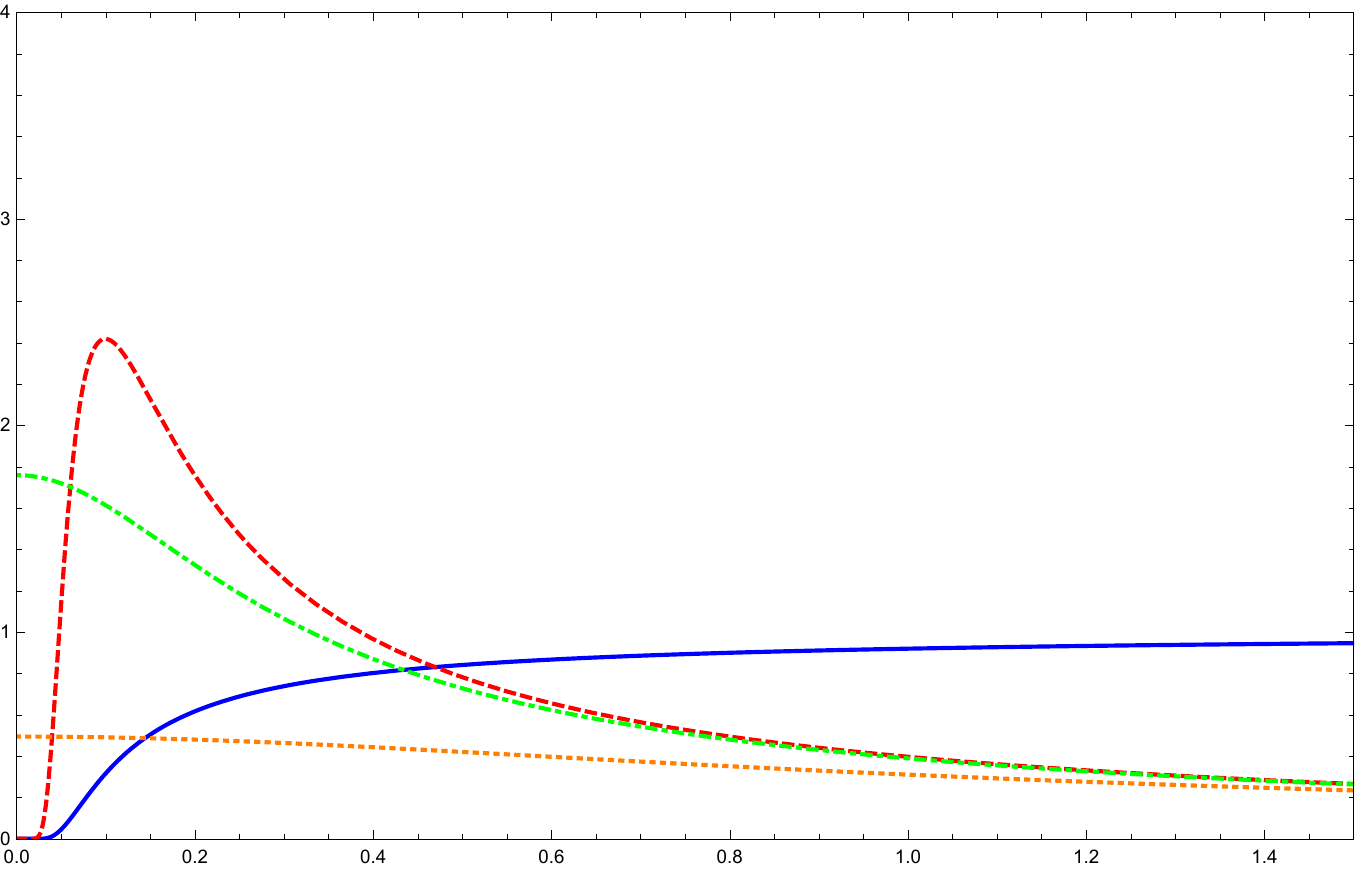}
	\captionsetup{justification=centering}
	\caption{weight functions w.r.t. $\sigma_x$ \\for $\mu_x-T=0.1$  }
\label{fig.weightsfunctionsse2}
			\end{center}
		\end{subfigure}
	\end{center}
	\vspace*{-3mm}
	\caption{Comparison of weight functions $w^{\mathrm{ls}}_T$ (blue, straight) and $w_{\sigma^2_\epsilon}$ for $\sigma^2_\epsilon=0.005$ (red, dashed),  $\sigma^2_\epsilon=0.2$ (green, dot-dashed) and  $\sigma^2_\epsilon=0.8$ (orange, dotted)  }
\end{figure}

 When $\sigma^2_\epsilon=0$,  $w_{\sigma^2_\epsilon}(x)$ is unbounded  for   $\sigma^2_x$   close to $0$ (see Fig. \ref{fig.weightsfunctionsmean}). This occurs for points located in the close neighborhood of an observation point and therefore, the future design points will be concentrated around already observed points.  At the opposite, when $\sigma^2_\epsilon$ is large,  weights tend to be uniform over $E$, resulting in a space-filling design. The choice of $\sigma^2_\epsilon$ will therefore influence the related optimal design.

  As shown in   Fig.\ref{fig.weightsfunctionsse2}, the main difference between the  weight functions  $w^{\mathrm{ls}}_T(x)$ and $w_{\sigma^2_\epsilon}(x)$ lies in   their behaviors  when  uncertainty is large. For a given value of  $|\mu_x-T|$ and a large value of $\sigma^2_x$, $w_{\sigma^2_\epsilon}(x)\approx 0$ whereas $w^{\mathrm{ls}}_T(x)\approx 1$.
   In section \ref{section.seqdesignexample}, we compare sequential designs build  w.r.t.   $w^{\mathrm{ls}}_T(x)$  and   $w_{\sigma^2_\epsilon}(x)$ by simulation studies. Note that the    weight function   $w^{\mathrm{ls}}_T(x)$ has no parameter   to   calibrate.

 \subsection{Areas exceeding a given threshold }
 \label{subsection.crit2}
 Here, we aim to build designs that give an accurate knowledge of the area $\{x\;;\; y(x) >T\}$ where the response  $y(x)$ exceeds a given threshold $T$.     In that case, we
 propose   the   weight function defined by

 \begin{equation}
 	\label{eq.weightpxexc}
 	w^{\mathrm{exc}}_T(x)=\mathbb{P}(y(x)>T)=F\left(\frac{\mu(x)-T}{\sigma_x}\right).
 \end{equation}

 We  give two interpretations of  the weight function $w^{\mathrm{exc}}_T(x)$ as above:
 $	w^{\mathrm{exc}}_T(x)$ is the p-value of the one-tailed hypothesis test $\mathcal H_0:``y(x)>T"$ vs $\mathcal H_1:``y(x)\le T"$. In the Bayesian interpretation with a flat prior on $y(x)$, $w^{\mathrm{exc}}_T(x)=\mathbb P(y(x)>T)$ is the posterior distribution of the event $"y(x)>T"$, given $\mu_x$. Therefore,   $w^{\mathrm{exc}}_T(x)$ corresponds to the faith that $x$ belongs to the target area.

 We denote by $\operatorname{MC}^{\mathrm{exc}}(d)$ and $\operatorname{IC}^{\mathrm{exc}}(d)$ the max-criterion (\ref{eq.Maxcrit}) and the integrated criterion (\ref{eq.integratedcrit})  related to $w^{\mathrm{exc}}_T(x)$.

 \section{Algorithms for optimal designs}
 \label{section.algorithm}
  We review some algorithms for obtaining optimal designs according to a generic criterion $C(d)$ that is to be minimized. In our applications, $C(d)$ will be either $\operatorname{MC}(d)=\max_{x\in E}c(x;d)$ or $\operatorname{IC}(d)=\sum_{x\in E}c(x;d)$, where $c(x;d)$ is defined by (\ref{eq.cxd}). We consider two types of design: non-sequential and sequential.

 Non-sequential designs rely entirely on prior information since all observation points are chosen before the experiment. They are suitable for parallelized computer experiments or sampling campaigns where responses are analyzed after the fact.

In adaptive sequential designs, on the other hand, responses on design points are observed after each stage. Therefore, the optimal designs points at a given stage can be based on the meta-model updated by the observations from the previous stages.

 \subsection{Non-sequential design}
 \label{Nseqdesignalgo}

The standard strategy for computationally obtaining an optimal design is to iteratively improve a fictitious starting design $d^{(0)}$ using an exchange algorithm (see  \cite{Jin2005} for a review).
Usually, the starting design $d^{(0)}$ is a standard space-filling or random design. However, these designs can be far from optimal and therefore require a large number of iterations to improve. We propose here a starting design, denoted $d^{\dag}$,   based on a cost-effective algorithm. Our simulation studies show that it is highly efficient (see figures \ref{districritinit} and \ref{districritinitls}).
 ~

 \subsubsection{Construction of the starting design  $d^{\dag}$}
 \label{secinitalgo}

 The starting design $d^{(0)}= d^\dag=\{x_1 ,x_2 ,\ldots,x_n\}$ is constructed from an empty design by sequentially adding one point at a time. Each new point maximizes (\ref{eq.cxd}) with updated variance and unchanged mean functions.
Note that this algorithm only requires the calculation of $n-1$   updated covariance matrices.

Start with the empty design $d_0=\emptyset$ and a given meta-model $\mathcal M_0: $ $\mathcal N(\mu,\Sigma)$.  Define the weight function $w_0(x)$, as in (\ref{eq.weightpxexc}), (\ref{eq.weightpxls}) or (\ref{eq.weightpicheny}) depending on the goal of the experiment.
Choose the point $x_1$ that maximizes $h_0(x)=w_0(x)\times \mathrm{Var}\left(y(x)|d_0\right)$ and set $d_1=\{x_1\}$.  Define the updated meta-model $\mathcal M_1$, with the same mean function $\mu(x)$ as the initial meta-model $\mathcal M_0$ and with the updated variance $\Sigma_1$ given by (\ref{updatedvar}) with $d=d_1$.  Note that, since the point $x_1$ is not observed, the mean function cannot be updated. Based on the new meta-model $\mathcal M_1$, we can define a new weight function $w_1(x)$.

 At the second step,    choose  $x_2$  that   maximizes  $w_1(x)\times \mathrm{Var}\left(y(x)|d_1\right)$.
 Put $d_2= \{x_1,x_2\}$, define a new meta-model $\mathcal M_2$ with the same mean function and the updated variance matrix given $d_2$. Then, define the new weight function $w_2(x)$, and so on... After $n$ iteration, we get the  $n$ points $\{x_1,...,x_n\}$ of $d^\dag$. This is detailed in Algorithm \ref{algo.d0star}.

 Note that the weight function $w_i(x)$  is based only on the mean function and the updated variance function  $\mathrm{var}(y(x)|d_{i-1})$. Therefore,  it is not necessary to compute the entire updated variance matrix,   only the diagonal terms.

 \begin{algorithm}[h]
 	\DontPrintSemicolon
 	\SetAlgoLined
 	\KwIn{Put $d_0=\emptyset$ the empty design ; \\ Initialize the prior meta-model as $\mathcal M_0 :$ $\mathcal N(\mu;\Sigma)$\\
 		Define the function  $w_0(x)$ from $\mathcal M_0$.
 	}
 	\ForEach{$i$ from $1$ to $n$}
 	{
 		\ForEach{$x$ in $E$}
 		{
 			Compute $h_i(x)=w_{i-1}(x)\times \mathrm{Var}\left(y(x)|d_{i-1}\right)$ ;\\
 		}
 		Choose $x_i=\mathrm{ArgMax}_{x\in E}\left(h_i(x)\right)$ ;\\
 		Put $d_i=\left\{x_i\right\}\cup d_{i-1}$ ;\\
 		Put  $\mathcal M_i:$  $\mathcal N(\mu,\Sigma_i)$, where $\Sigma_i$ is the updated covariance w.r.t. $d_i$ ($\mu$ is unchanged).\\
 		Define the function  $w_i(x)$ from $\mathcal M_i$.
 	}
 	\KwOut{design $d^{\dag}=d_n$.}
 	\caption{Construction of the  n-point design $d^{\dag}$.}
 	\label{algo.d0star}
 \end{algorithm}

 \subsubsection{Exchange algorithm}
 \label{secexchalgo}

 From a starting design $d^{(0)}$, the exchange algorithm, see \cite{Kennard1969,Mitchell1974}, displayed in Algorithm \ref{algo.exchange},
 consists of iterative permutations of the design's inner and outer points. A permutation is accepted if the corresponding criterion is improved.  We denote $d^{(k)}$ the design obtained after $k$ iterations.

 \begin{algorithm}[h]
 	
 	\DontPrintSemicolon
 	\SetAlgoLined
 	\KwIn{Choose a starting design $d^{(0)}$, maximal number of iterations $M$ ;}
 	\ForEach{$k$ from $1$ to $M$}
 	{
 		Randomly draw $x\in d^{(k-1)}$ ;\\
 		Randomly draw $x'\in \bar{d}^{(k-1)}$ ;\\
 		Permute $x$ and $x'$ considering $d_{temp}=d^{(k-1)}\cup\{x'\}\backslash\{x\}$ ;\\
 		\eIf{$C\left(d_{temp}\right)<C\left(d^{(k-1)}\right)$}
 		{
 			$d^{(k)}=d_{temp}$;
 		}
 		{
 			$d^{(k)}=d^{(k-1)}$;
 		}
 	}
 	\KwOut{design $d^*=d^{(M)}$.}
 	\caption{Exchange algorithm.}
 	\label{algo.exchange}
 \end{algorithm}

In addition, simulated annealing \cite{Morris1995} can be used within the exchange algorithm to avoid local extrema issues. However, we did not observe any improvement over the exchange algorithm in our simulation studies.

 \subsection{Sequential designs}
 \label{subsection.seqdeisgnalgo}

 A sequential design is divided in  $N$ stages. At   Stage $i$, we have to build a   $n_i$-point design $d_i$ based on the updated meta-model, which consists of an initial meta-model updated with the observations obtained in the previous stages (see \cite{Williams2000,Bect2012,Villanueva2016}).
 The new design $d_i$ can be constructed by using Algorithm \ref{algo.d0star} and \ref{algo.exchange}  with the updated meta-model obtained with  Formulas (\ref{updatedmean}) and (\ref{updatedvar}).    Examples of sequential designs are given in  Section \ref{section.seqdesignexample}.

 \subsection{Efficiency factor}
 \label{section.efficiency}
 For any  $n$-point  design $d$, the  theoretical efficiency factor of $d$ is defined by
 $$	\textrm{eff}(d)=\frac{C(d^{opt})}{C(d)}.
 $$
 where $d^{opt}$ is the  theoretical $n$-point design that  minimizes the criterion $C(d)$. Since  $d^{opt}$  is not   attainable, we approximate eff(d) by

 \begin{equation}
 	\label{eq.efficiency}
 	\textrm{eff}(d)=\frac{C(d^{**})}{C(d)}.
 \end{equation}
 where $d^{**}$ is computed as follow: we draw 1000 $n$-point   designs $(d^{(0)}_1,..., d^{(0)}_{1000})$, where the $n$ points are chosen at random over the grid. For each design $d_i^{(0)}$, we apply the exchange algorithm and obtain a design $d_i^{*}$. The design $d^{**}$ is the design that minimizes $C(d)$ among $(d_1^{*}, ... d_{1000}^{*}).$

 \section{Simulation studies for sequential designs}
 \label{section.seqdesignexample}
 In this section, the aim of the experiment is to provide an accurate estimation of the level set  $\mathcal{L}=\{x\,:\,y(x)=T\} $ for a given $T$ on a $150\times 150$ grid.
 We consider sequential adaptive designs with one additional point at each stage.  Once the design point is observed,  the level set is estimated by  $\widehat{\mathcal L}=\{x\in E\;;\;\widehat{y}(x)=T\}$ where $\widehat{y}(x)$ is the mean of the updated meta-model.

 Before starting the experiments, we assume no prior information on the mean. In this case,  we recommend to choose,  for all  $x\in E$,  $\mu(x)=T$   as a default choice for the mean function of the meta-model (\ref{eq.metamodel}). This choice implies that any point $x$  potentially belongs to the level set $\mathcal{L}$. For the  covariance matrix   of the meta-model,
 we assume that  $\Sigma$  the Mat\'ern covariance function \begin{equation}\mathrm{cov}(x_i,x_j)=2^{1-\nu}\sigma^2(d\sqrt{2\nu}/\kappa)^\nu K_\nu(d\sqrt{2\nu}/\kappa)/\Gamma(\nu)
 \end{equation}
 with  $\sigma=0.7$, $\nu=0.7$ and $\kappa=0.2$,  where  $d=d(x_i,x_j)$ is the Euclidean distance between $x_i$ and $ x_j$, $\Gamma$ is the Gamma function and $K$ the modified Bessel function.

 As we have no prior information on the mean,  we start with a   space-filling design with few points. Once the design points are observed, we update the mean and the variance of the meta-model by applying Formula \ref{updatedmean} and \ref{updatedvar}.
 Then,  we  start the sequential design with one of our criteria, $\operatorname{IC}^{\mathrm{ls}}$
 or
 $\operatorname{MC}^{\mathrm{ls}}$,
 based on the updated  meta-model. At each stage, we construct the 1-point design that minimizes the updated criterion.

 We compare our designs with those obtained by minimizing the criteria $\operatorname{IC}^{W}$     or $\operatorname{MC}^{W}$. To calibrate the parameter $\sigma^2_\varepsilon$, we follows the recommendation given in   \cite{Picheny2010}, i.e. we choose  choose  $\sigma^2_\varepsilon=(\max_x \mu(x)-\min_x\mu(x))/20$ which is updated at  each stage.

To compare the relative performance of the models in terms of level set estimation accuracy, we propose three quality scores below:

 \begin{figure}
 	\begin{center}
 		\begin{subfigure}[b]{0.48\textwidth}
 			\begin{center}
 				\includegraphics[scale=0.181,trim={30 30 10 60},clip]{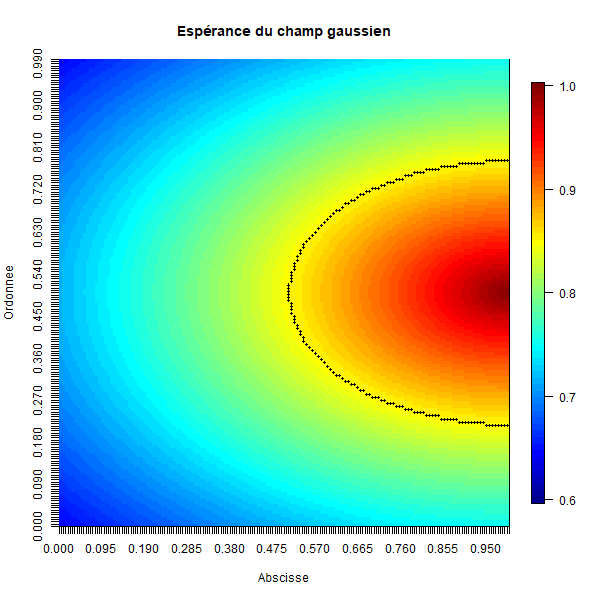}
 				\captionsetup{justification=centering}
 				\caption{Example 1:\\ smooth   level set.}
 				\label{ex.ls.smooth}
 			\end{center}
 		\end{subfigure}
 		\begin{subfigure}[b]{0.48\textwidth}
 			\begin{center}
 				\includegraphics[scale=0.181,trim={30 30 10 60},clip]{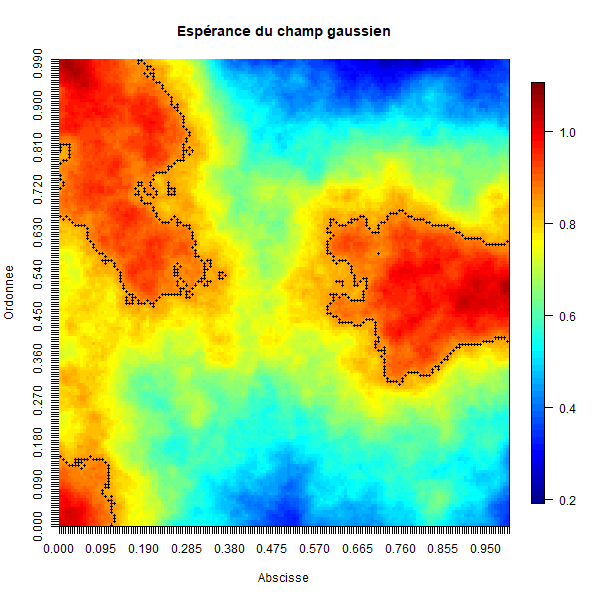}
 				\captionsetup{justification=centering}
 				\caption{Example 2: \\complex disconnected level set.}
 				\label{ex.ls.complex}
 			\end{center}
 		\end{subfigure}
 	\end{center}
 	\vspace*{-3mm}
 	\caption{Map of $y(x)$ and level set $\mathcal L=\{x\,:\,y(x)=0.85\}$ (black line). }
 \end{figure}

 \subsubsection*{Score $Q_\mathrm{dist}$: distance average}
\begin{figure}
 	\captionsetup{justification=centering}
 	\begin{center}
 		\begin{subfigure}[b]{0.45\textwidth}
 			\begin{center}
 				\includegraphics[scale=0.12]{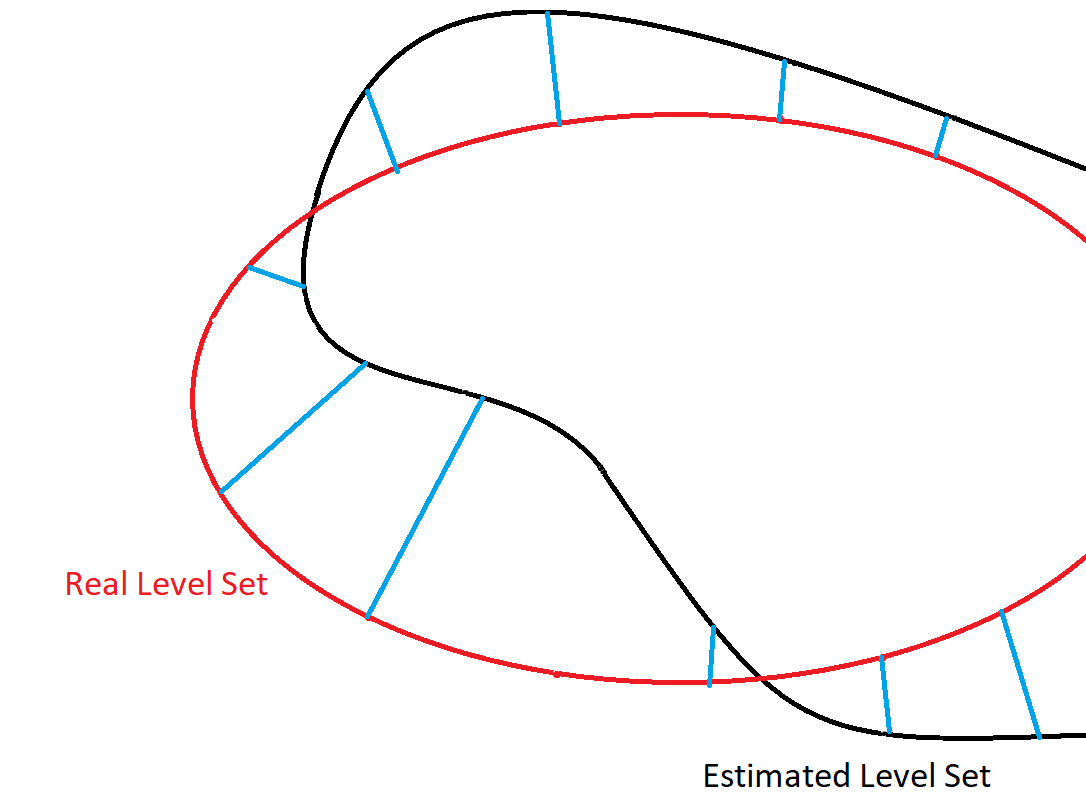}
 				\captionsetup{justification=centering}
 				\caption{$Q_\mathrm{dist}^\mathrm{actual}$ score }
 				\label{fig.scoredist}
 			\end{center}
 		\end{subfigure}
 		\begin{subfigure}[b]{0.45\textwidth}
 			\begin{center}
 				\includegraphics[scale=0.12]{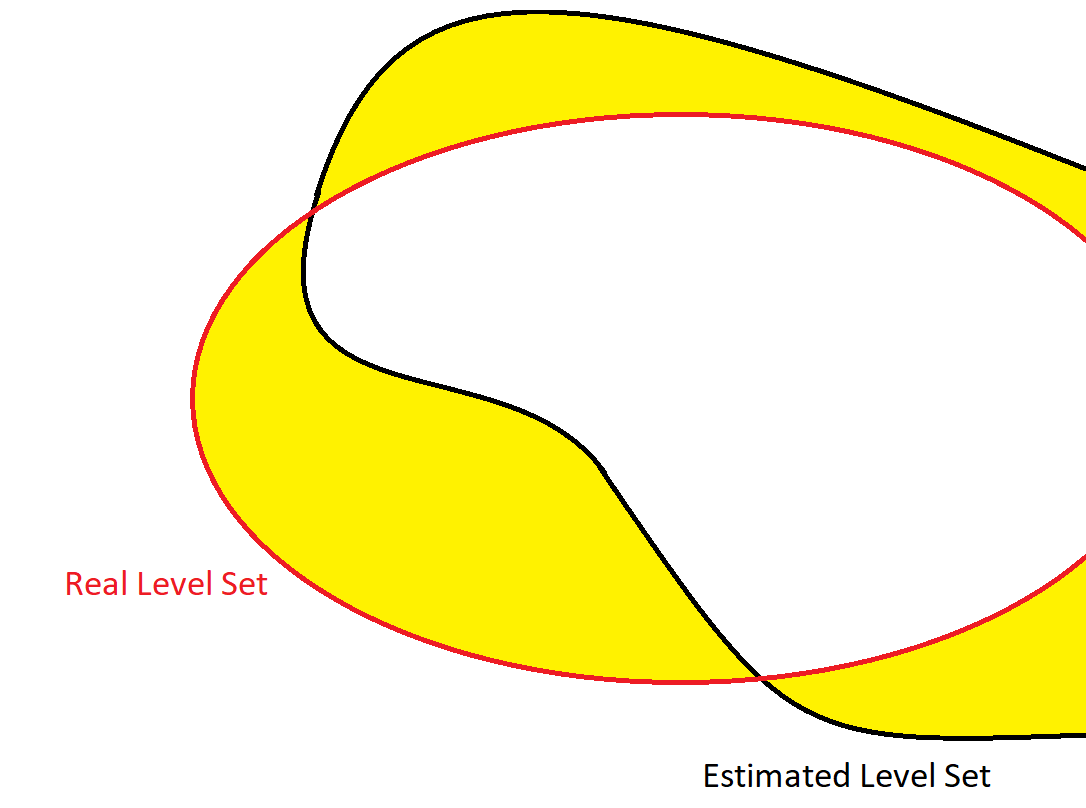}
 				\captionsetup{justification=centering}
 				\caption{$Q_\mathrm{area}$ score}
 				\label{fig.scorearea}
 			\end{center}
 		\end{subfigure}
 	\end{center}
 	\vspace*{-3mm}
 	\caption{Quality scores for estimated level set}
 \end{figure}

 This score  evaluates   the symmetric distance between the estimated and actual level sets. It is defined by :

 $$Q_\mathrm{dist}=\frac{1}{2}\left(Q^{\mathrm{actual}}_\mathrm{dist}+Q^{\mathrm{est}}_\mathrm{dist}\right),$$
 where
 $$Q^{\mathrm{actual}}_\mathrm{dist}=\frac{1}{\#\{\mathcal{L}\}}\sum_{x\in\mathcal{L}}\mathrm{min}_{x'\in\widehat{\mathcal L}}\;d(x,x'),$$
 is the average distance between each point of the actual level set and the nearest point in  the estimated level set (see Fig. \ref{fig.scoredist}). Symmetrically,

 $$Q^{\mathrm{est}}_\mathrm{dist}=\frac{1}{\#\{\widehat{\mathcal L}\}}\sum_{x\in\widehat{\mathcal L}}\mathrm{min}_{x'\in\mathcal{L}}\;d(x,x'),$$
is the average distance between each point in the estimated level set and the nearest point in the actual level set.

 \subsubsection*{Score  $Q_\mathrm{value}$: values average}

 This score evaluates the average discrepancy between the  estimated and actual values on both the actual and estimated level sets:

 $$Q_\mathrm{value}=\frac{1}{2}\left(Q^{\mathrm{actual}}_\mathrm{value}+Q^{\mathrm{est}}_\mathrm{value}\right), $$
 where
 $$Q^{\mathrm{est}}_\mathrm{value}=\frac{1}{\#\{\widehat{\mathcal L}\}}\sum_{x\in\widehat{\mathcal L}}|y(x)-T|,$$
 and
 $$Q^{\mathrm{actual}}_\mathrm{value}=\frac{1}{\#\{\mathcal{L}\}}\sum_{x\in\mathcal{L}}|\widehat y(x)-T|.$$

 \subsubsection*{Score $Q_\mathrm{area}$:  area   between the two level sets}

 This score corresponds to the proportion of area between the actual level set $\mathcal L$ and the estimated one $\widehat{\mathcal L}$, (see Fig. \ref{fig.scorearea}): $$Q_\mathrm{area}=\frac{1}{N^2}\;\#\Big\{x\in E\;:\;\big(\widehat{y}(x)<T\textrm{ and }y(x)>T\big)\textrm{ or }\big( \widehat{y}(x)>T\textrm{ and } y(x)<T\big)\Big\}.$$

 It can also be seen as the proportion of points misclassified in the sets  $\{x:y(x)>T\}$ and  $\{x:y(x)<T\}$.

 \subsubsection*{Other scores}
 Both $Q_\mathrm{dist}$ and $Q_\mathrm{value}$ scores are defined as symmetrized average distances. The individual components   $Q^{\mathrm{est}}_\mathrm{value}$, $Q^{\mathrm{actual}}_\mathrm{value}$,  $Q^{\mathrm{est}}_\mathrm{dist}$ and $Q^{\mathrm{actual}}_\mathrm{dist}$   and their maximal alternatives have also been tested as quality scores. For smooth one-piece level sets, they give similar results to those obtained with $Q_\mathrm{dist}$ and $Q_\mathrm{value}$. When the target level set is disconnected, they are difficult to interpret. So,   they are not considered in this paper.

 \subsection{Example 1: smooth level set}
 \label{ex.seq.smooth}

\begin{figure}
 	\begin{center}
 		\begin{subfigure}[b]{0.242\textwidth}
 			\begin{center}
 				\includegraphics[scale=0.151,trim={30 30 10 60},clip]{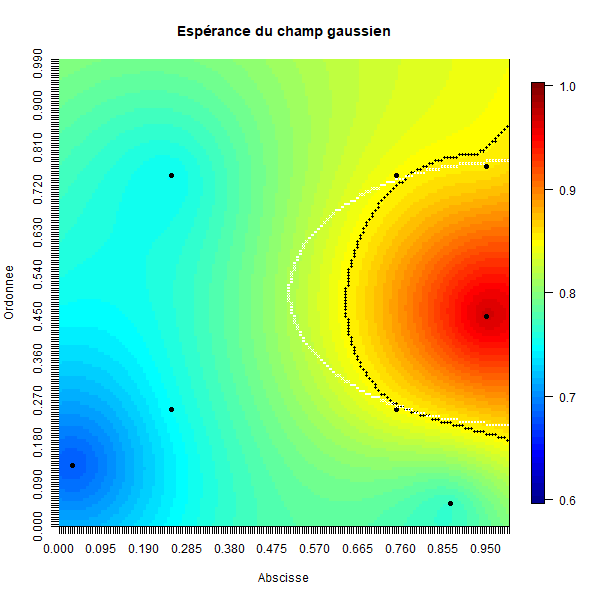}
 				\caption{Map for $\operatorname{MC}^{\mathrm{ls}}$.}
 			\end{center}
 		\end{subfigure}
 		\begin{subfigure}[b]{0.242\textwidth}
 			\begin{center}
 				\includegraphics[scale=0.151,trim={30 30 10 60},clip]{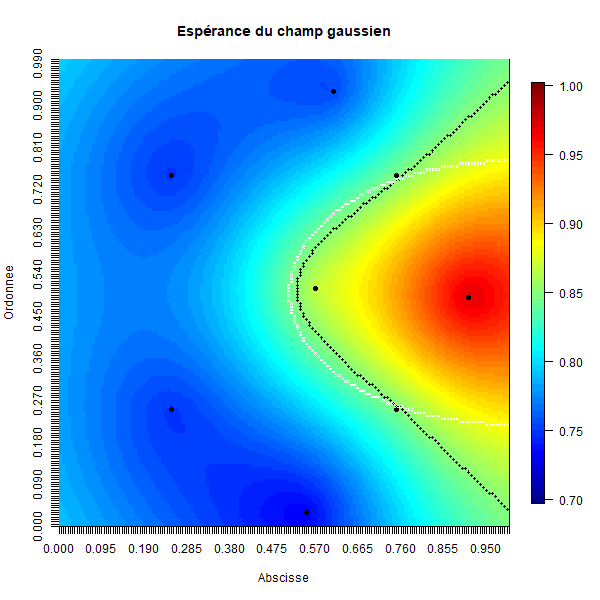}
 				\caption{Map for  $\operatorname{IC}^{\mathrm{ls}}$.}
 			\end{center}
 		\end{subfigure}
 		\begin{subfigure}[b]{0.242\textwidth}
 			\begin{center}
 				\includegraphics[scale=0.151,trim={30 30 10 60},clip]{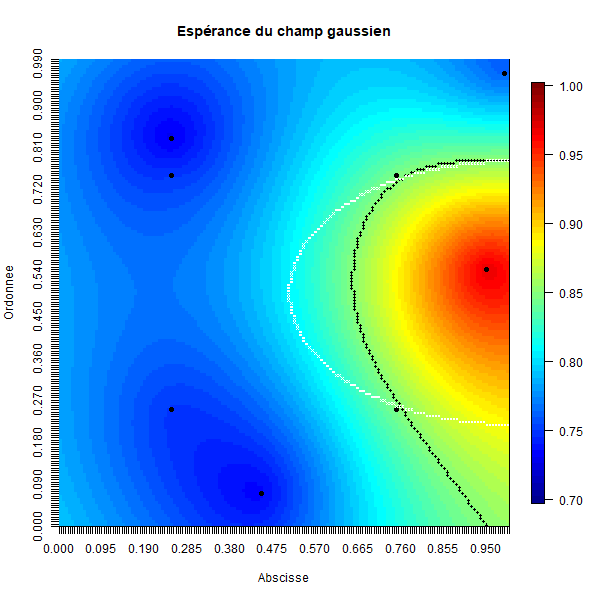}
 				\caption{Map for $\operatorname{MC}^{W}$.}
 			\end{center}
 		\end{subfigure}
 		\begin{subfigure}[b]{0.242\textwidth}
 			\begin{center}
 				\includegraphics[scale=0.151,trim={30 30 10 60},clip]{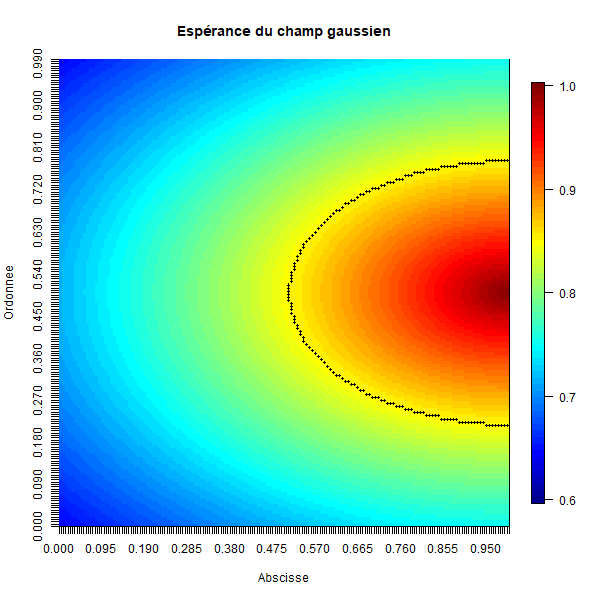}
 				\caption{Map for $\operatorname{IC}^{W}$.}
 			\end{center}
 		\end{subfigure}
 	\end{center}
 	\vspace*{-3mm}
 	
 	\caption{comparison   of    $\widehat y(x)$  maps   for $8$-point designs  (black dots) for Example 1. The actual (white)  and estimated (black) level sets are displayed.}
 	\label{fig.smooth.8points}
 \end{figure}

 We consider a smooth field  displayed in Fig. \ref{ex.ls.smooth} with the smooth level set $ \mathcal L=\left\{x\,:\,y(x)=T\right\}$. In this example, we choose
 $T=0.85$. First, we perform a 4-point space-filling minimax design to acquire information.  Then, for  each  criterion
 $\operatorname{MC}^{\mathrm{ls}}$, $\operatorname{IC}^{\mathrm{ls}}$, $\operatorname{MC}^{W}$ and $\operatorname{IC}^{W}$,  we construct the  sequential design. The estimated level sets $\widehat{ \mathcal L}=\left\{x\,:\,\widehat y(x)=T\right\}$ are displayed in Fig. \ref{fig.smooth.8points} for 8-point designs, which correspond to four stages after the starting space-filling design.

\begin{figure}
 	\begin{center}
 		\begin{subfigure}[b]{0.32\textwidth}
 			\begin{center}
 				\includegraphics[scale=0.24,trim={30 30 10 40},clip]{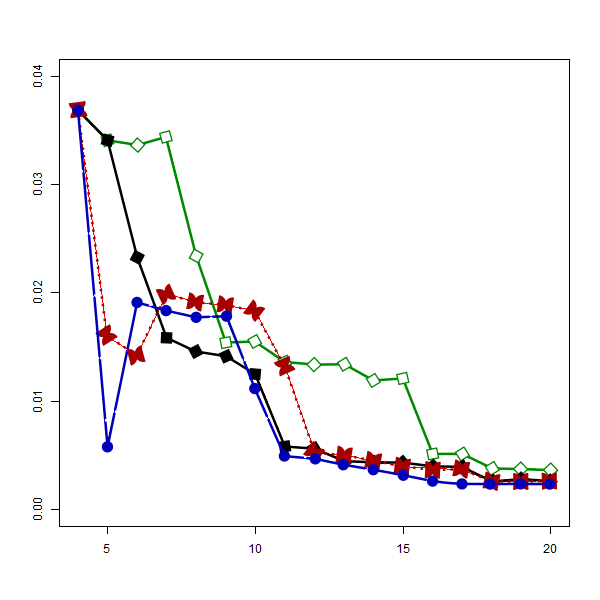}
 				\caption{  $Q_\mathrm{value}$   score}
 			\end{center}
 		\end{subfigure}
 		\begin{subfigure}[b]{0.32\textwidth}
 			\begin{center}
 				\includegraphics[scale=0.24,trim={30 30 10 40},clip]{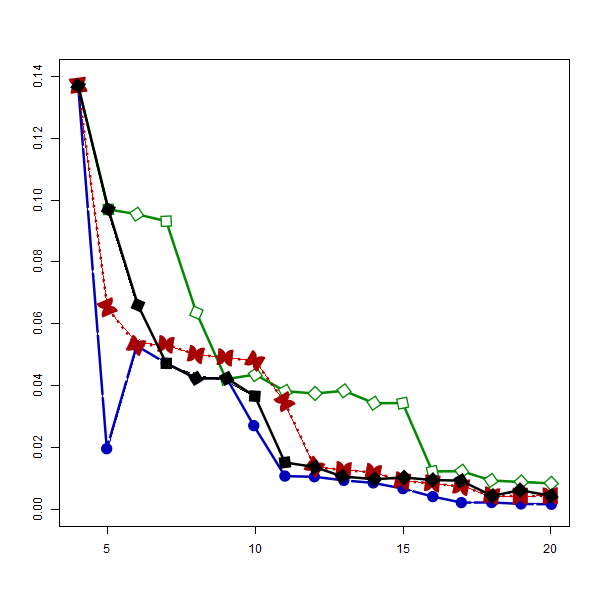}
 				\caption{  $Q_\mathrm{dist}$   score}
 			\end{center}
 		\end{subfigure}
 		\begin{subfigure}[b]{0.32\textwidth}
 			\begin{center}
 				\includegraphics[scale=0.24,trim={30 30 10 40},clip]{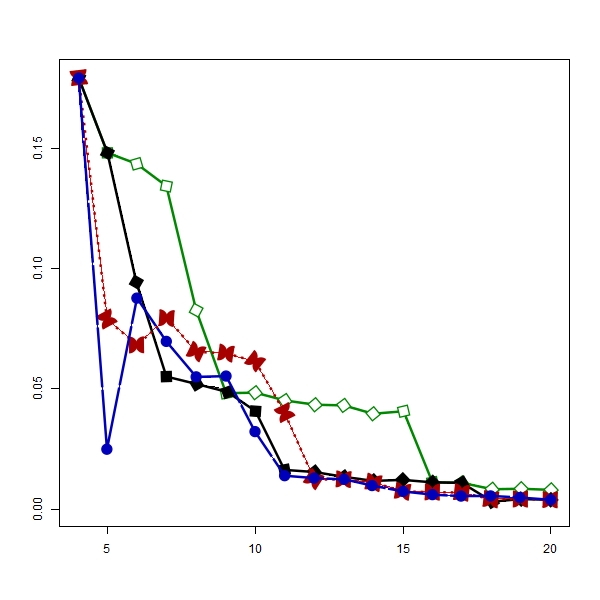}
 				\caption{ $Q_\mathrm{area}$   score}
 			\end{center}
 		\end{subfigure}
 	\end{center}
 	\caption{quality scores against  the number of design points    w.r.t criteria $\operatorname{MC}^{\mathrm{ls}}$ (black), $\operatorname{IC}^{\mathrm{ls}}$ (blue), $\operatorname{IC}^{W}$ (red) and $\operatorname{MC}^{W}$ (green) for Example 1.}
 	\label{fig.score.smooth}
 \end{figure}

 In  Fig. \ref{fig.score.smooth}, we compare the performance of the four strategies against the three quality scores proposed above. There is no significant difference between the criteria, except with $\operatorname{MC}^W$ which seems to be globally less efficient.   After 12 stages,  all the 16-point designs have the same performance.

 \subsection{Example 2: complex level set}
 \label{ex.seq.complex}

 In this example, we consider a more complex field displayed in \ref{ex.ls.complex}. The level set to be estimated has three irregular disconnect components. First, we perform a 3-point minimax design.  Then, for  each  criterion
 $\operatorname{MC}^{\mathrm{ls}}$, $\operatorname{IC}^{\mathrm{ls}}$, $\operatorname{MC}^{W}$ and $\operatorname{IC}^{W}$, we construct a sequential design with one additional point at each stage.

\begin{figure}
 	\begin{center}
 		6-point designs\vspace*{5pt}

 		\begin{subfigure}[b]{0.242\textwidth}
 			\begin{center}
 				\includegraphics[scale=0.151,trim={30 30 10 60},clip]{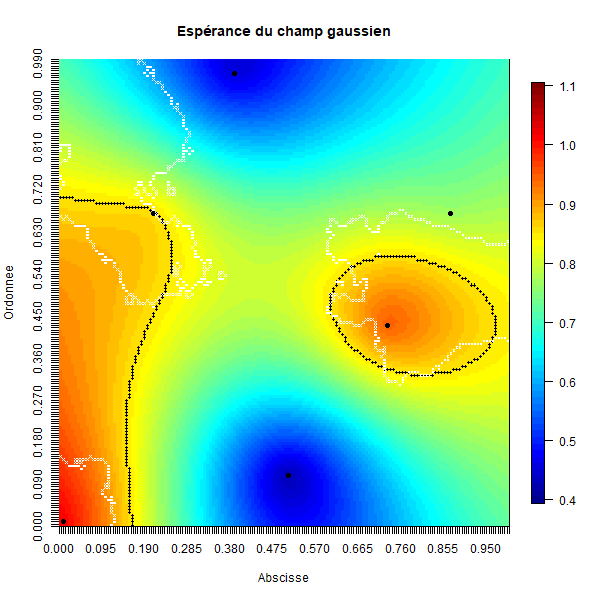}
 				\caption{Map for $\operatorname{MC}^{\mathrm{ls}}$.}
 			\end{center}
 		\end{subfigure}
 		\begin{subfigure}[b]{0.242\textwidth}
 			\begin{center}
 				\includegraphics[scale=0.151,trim={30 30 10 60},clip]{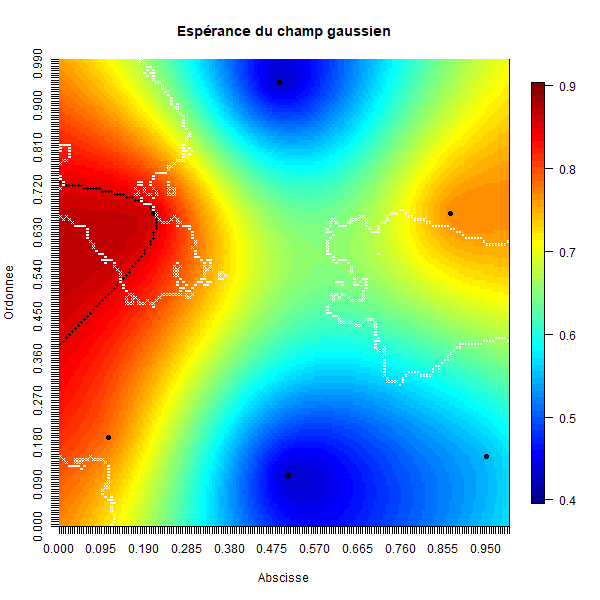}
 				\caption{Map for $\operatorname{IC}^{\mathrm{ls}}$.}
 			\end{center}
 		\end{subfigure}
 		\begin{subfigure}[b]{0.242\textwidth}
 			\begin{center}
 				\includegraphics[scale=0.151,trim={30 30 10 60},clip]{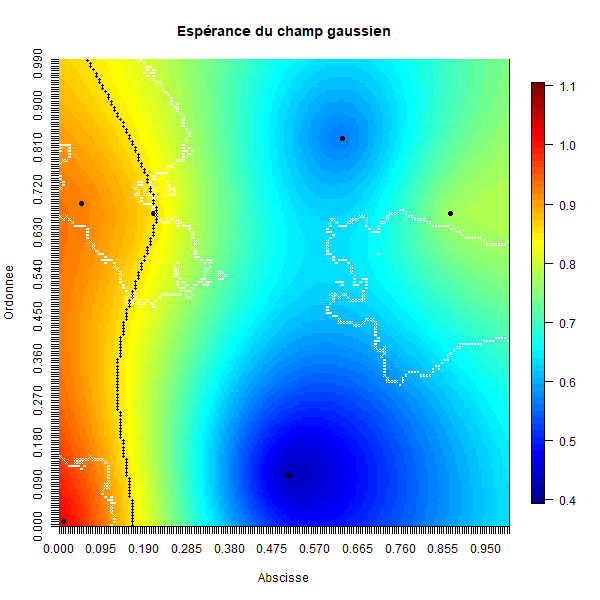}
 				\caption{Map for $\operatorname{MC}^{W}$.}
 			\end{center}
 		\end{subfigure}
 		\begin{subfigure}[b]{0.242\textwidth}
 			\begin{center}
 				\includegraphics[scale=0.151,trim={30 30 10 60},clip]{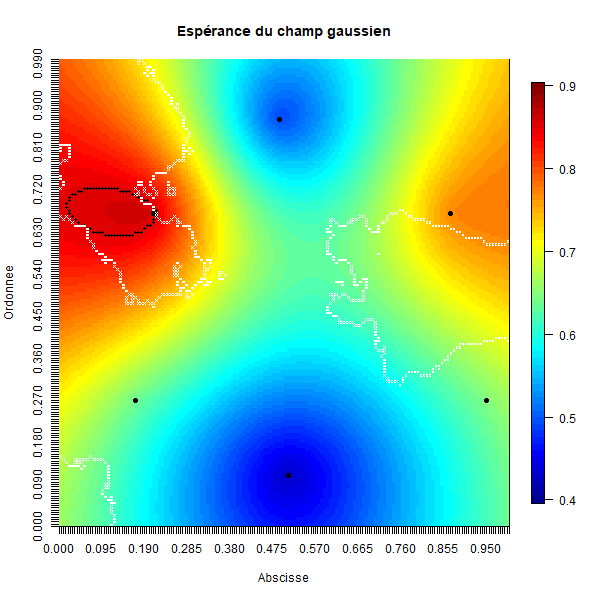}
 				\caption{Map for $\operatorname{IC}^{W}$.}
 			\end{center}
 		\end{subfigure}
 	\end{center}

 	\begin{center}
 		
 		10-point designs \vspace*{5pt}
 		
 		\begin{subfigure}[b]{0.242\textwidth}
 			\begin{center}
 				\includegraphics[scale=0.151,trim={30 30 10 60},clip]{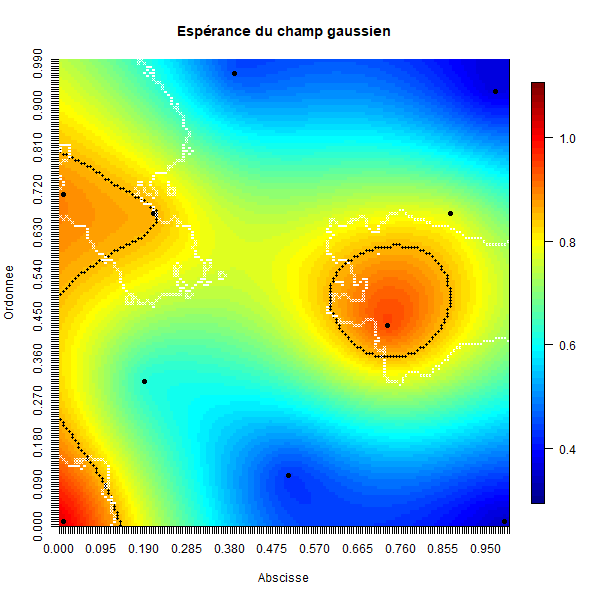}
 				\caption{Map for $\operatorname{MC}^{\mathrm{ls}}$.}
 			\end{center}
 		\end{subfigure}
 		\begin{subfigure}[b]{0.242\textwidth}
 			\begin{center}
 				\includegraphics[scale=0.151,trim={30 30 10 60},clip]{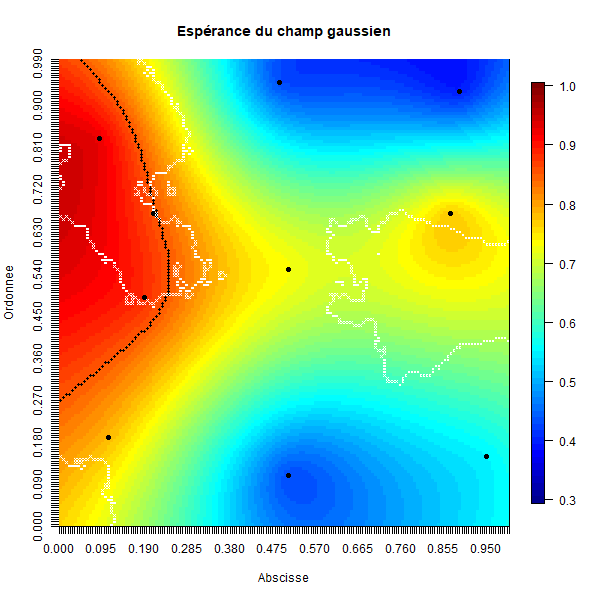}
 				\caption{Map for $\operatorname{IC}^{\mathrm{ls}}$.}
 			\end{center}
 		\end{subfigure}
 		\begin{subfigure}[b]{0.242\textwidth}
 			\begin{center}
 				\includegraphics[scale=0.151,trim={30 30 10 60},clip]{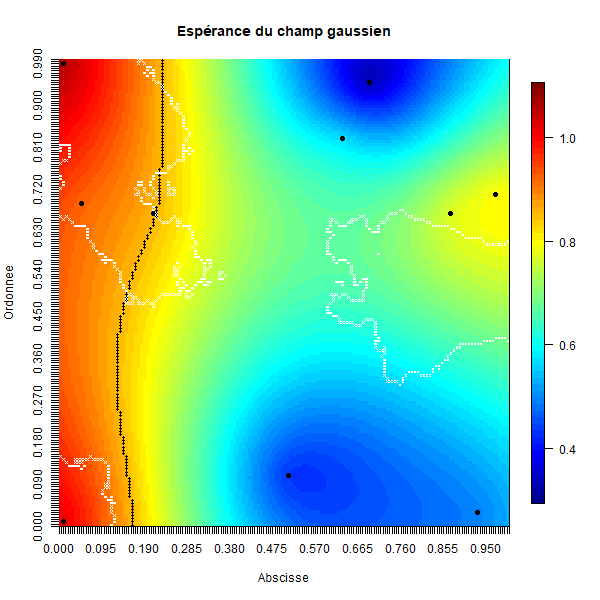}
 				\caption{Map for $\operatorname{MC}^{W}$.}
 			\end{center}
 		\end{subfigure}
 		\begin{subfigure}[b]{0.242\textwidth}
 			\begin{center}
 				\includegraphics[scale=0.151,trim={30 30 10 60},clip]{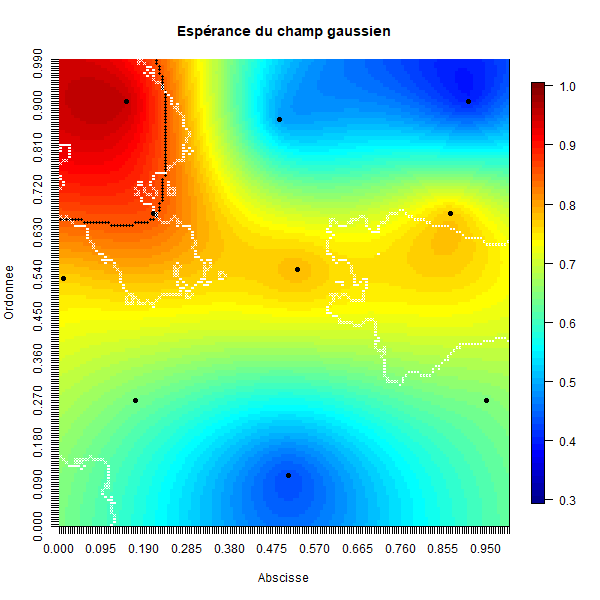}
 				\caption{Map for $\operatorname{IC}^{W}$.}
 			\end{center}
 		\end{subfigure}
 	\end{center}
 	
 	\begin{center}
 		
 		13-point designs \vspace*{5pt}

 		\begin{subfigure}[b]{0.242\textwidth}
 			\begin{center}
 				\includegraphics[scale=0.151,trim={30 30 10 60},clip]{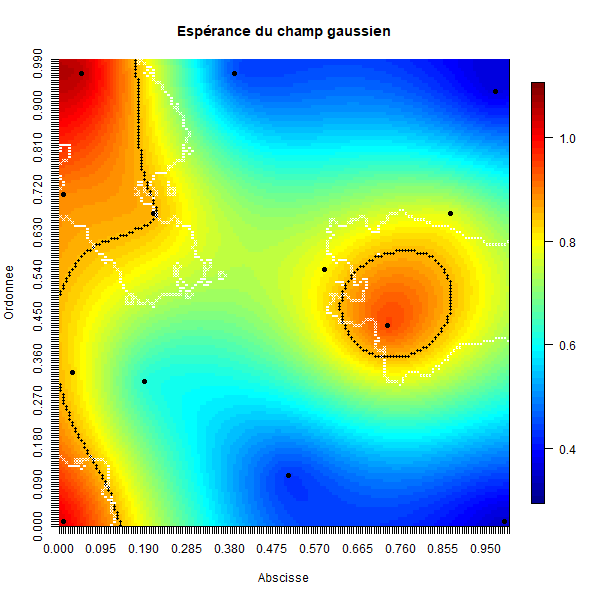}
 				\caption{Map for $\operatorname{MC}^{\mathrm{ls}}$.}
 			\end{center}
 		\end{subfigure}
 		\begin{subfigure}[b]{0.242\textwidth}
 			\begin{center}
 				\includegraphics[scale=0.151,trim={30 30 10 60},clip]{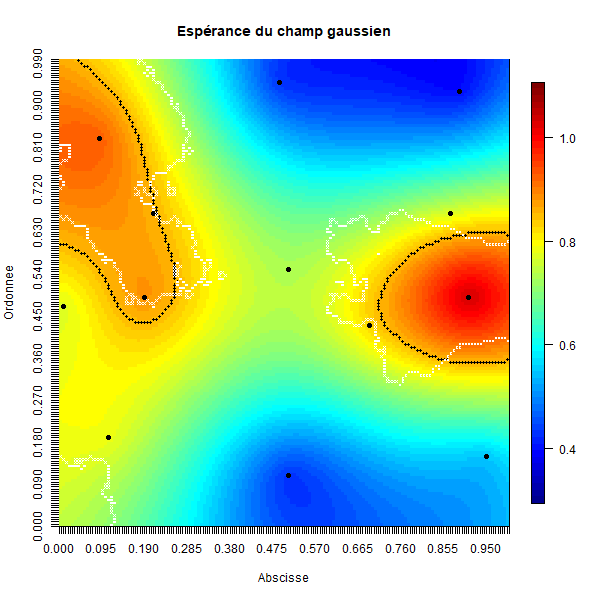}
 				\caption{Map for $\operatorname{IC}^{\mathrm{ls}}$.}
 			\end{center}
 		\end{subfigure}
 		\begin{subfigure}[b]{0.242\textwidth}
 			\begin{center}
 				\includegraphics[scale=0.151,trim={30 30 10 60},clip]{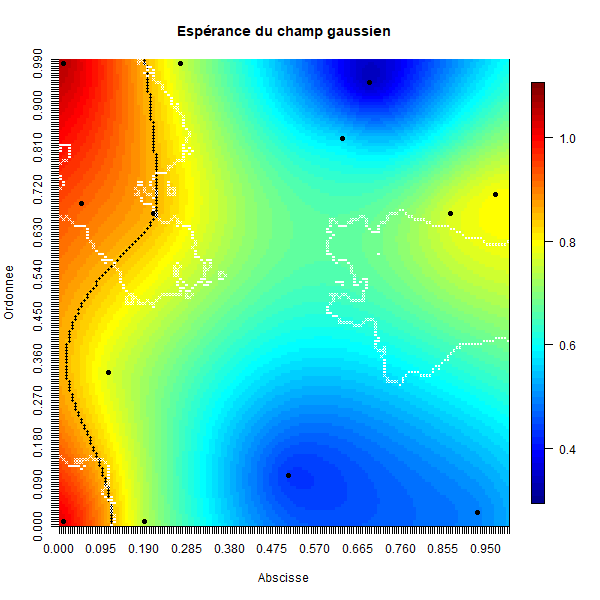}
 				\caption{Map for $\operatorname{MC}^{W}$.}
 			\end{center}
 		\end{subfigure}
 		\begin{subfigure}[b]{0.242\textwidth}
 			\begin{center}
 				\includegraphics[scale=0.151,trim={30 30 10 60},clip]{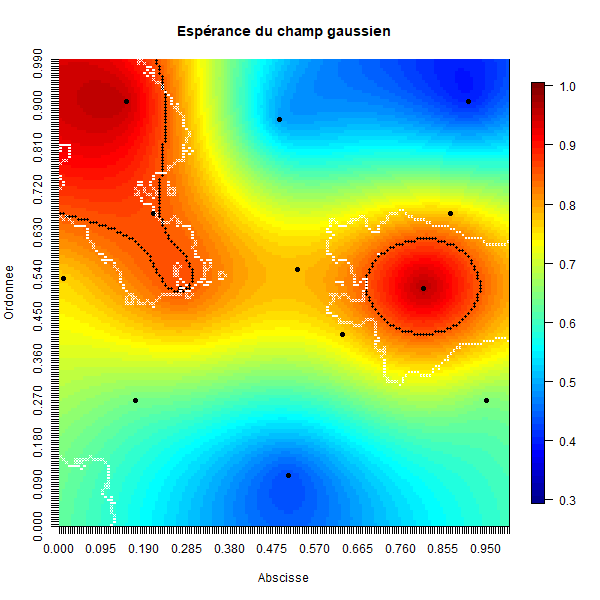}
 				\caption{Map for $\operatorname{IC}^{W}$.}
 			\end{center}
 		\end{subfigure}

 		17-point designs \vspace*{5pt}

 		\begin{subfigure}[b]{0.242\textwidth}
 			\begin{center}
 				\includegraphics[scale=0.151,trim={30 30 10 60},clip]{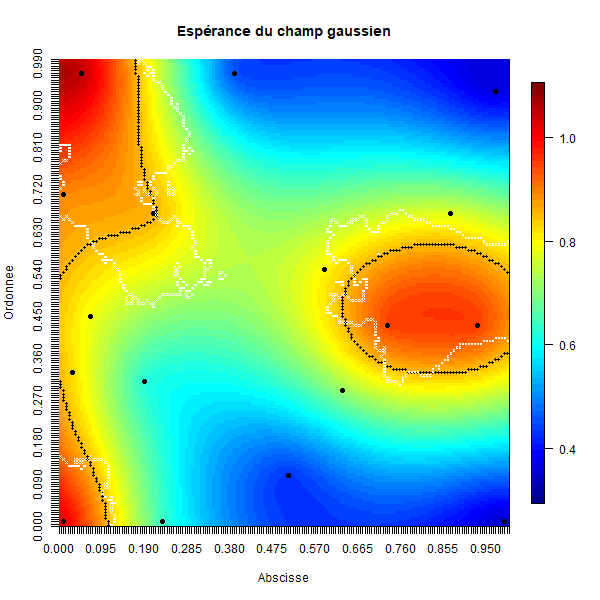}
 				\caption{Map for $\operatorname{MC}^{\mathrm{ls}}$.}
 			\end{center}
 		\end{subfigure}
 		\begin{subfigure}[b]{0.242\textwidth}
 			\begin{center}
 				\includegraphics[scale=0.151,trim={30 30 10 60},clip]{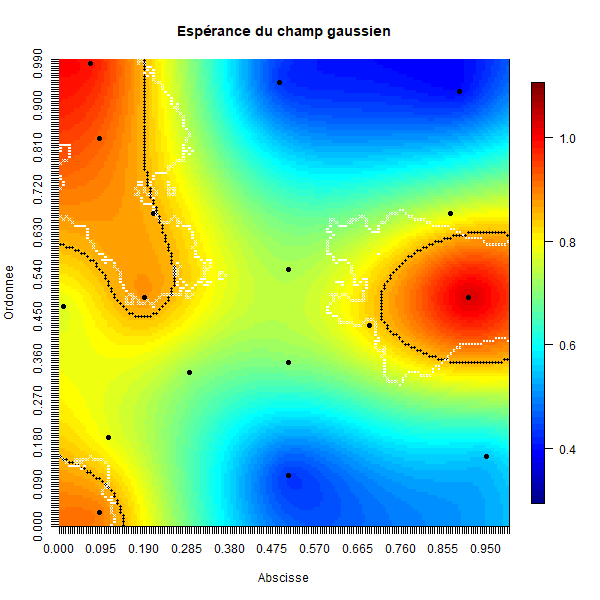}
 				\caption{Map for $\operatorname{IC}^{\mathrm{ls}}$.}
 			\end{center}
 		\end{subfigure}
 		\begin{subfigure}[b]{0.242\textwidth}
 			\begin{center}
 				\includegraphics[scale=0.151,trim={30 30 10 60},clip]{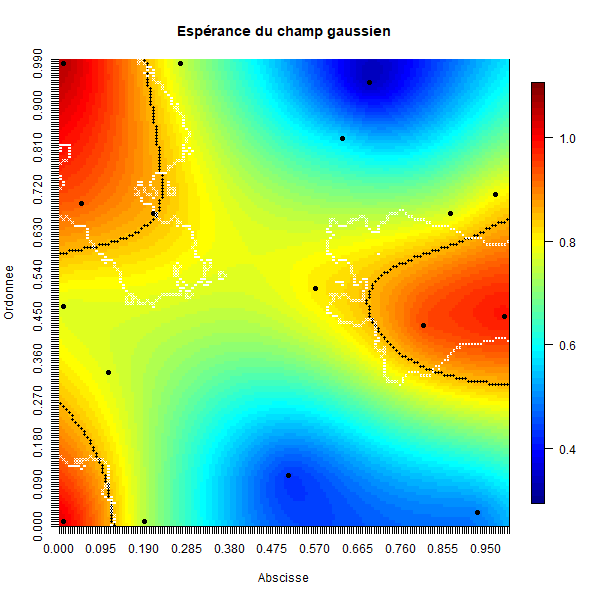}
 				\caption{Map for $\operatorname{MC}^{W}$.}
 			\end{center}
 		\end{subfigure}
 		\begin{subfigure}[b]{0.242\textwidth}
 			\begin{center}
 				\includegraphics[scale=0.151,trim={30 30 10 60},clip]{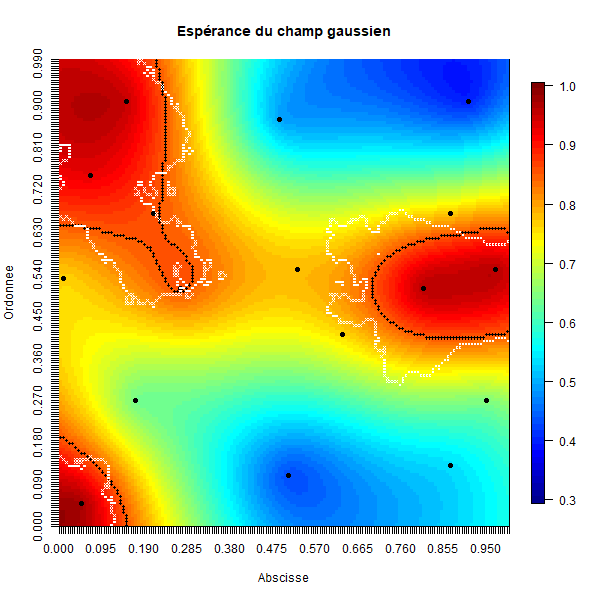}
 				\caption{Map for $\operatorname{IC}^{W}$.}
 			\end{center}
 		\end{subfigure}
 		
 		20-point designs \vspace*{5pt}

 		\begin{subfigure}[b]{0.242\textwidth}
 			\begin{center}
 				\includegraphics[scale=0.151,trim={30 30 10 60},clip]{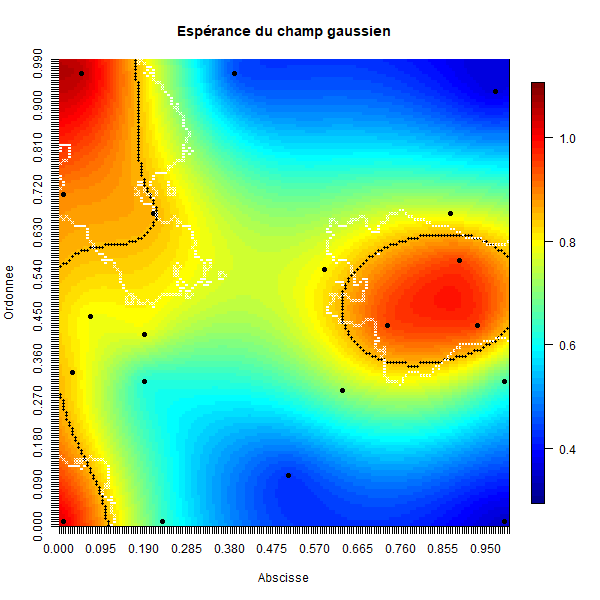}
 				\caption{Map for $\operatorname{MC}^{\mathrm{ls}}$.}
 			\end{center}
 		\end{subfigure}
 		\begin{subfigure}[b]{0.242\textwidth}
 			\begin{center}
 				\includegraphics[scale=0.151,trim={30 30 10 60},clip]{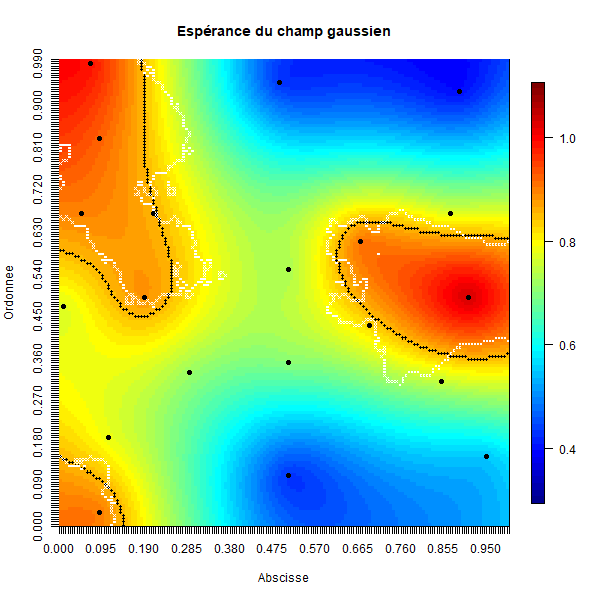}
 				\caption{Map for $\operatorname{IC}^{\mathrm{ls}}$.}
 			\end{center}
 		\end{subfigure}
 		\begin{subfigure}[b]{0.242\textwidth}
 			\begin{center}
 				\includegraphics[scale=0.151,trim={30 30 10 60},clip]{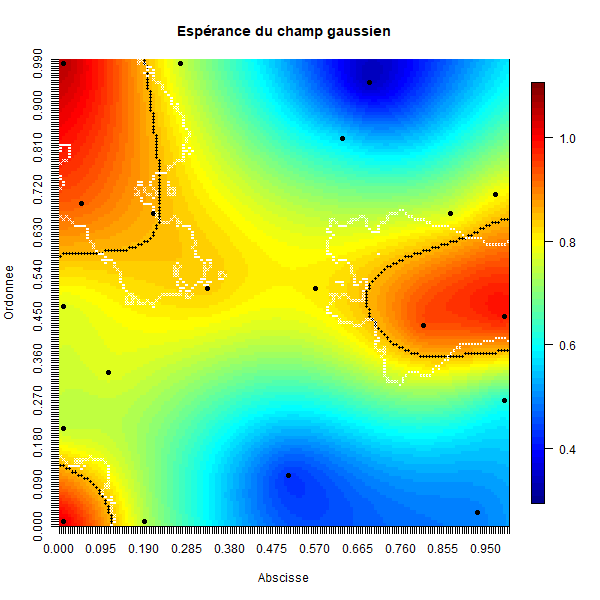}
 				\caption{Map for $\operatorname{MC}^{W}$.}
 			\end{center}
 		\end{subfigure}
 		\begin{subfigure}[b]{0.242\textwidth}
 			\begin{center}
 				\includegraphics[scale=0.151,trim={30 30 10 60},clip]{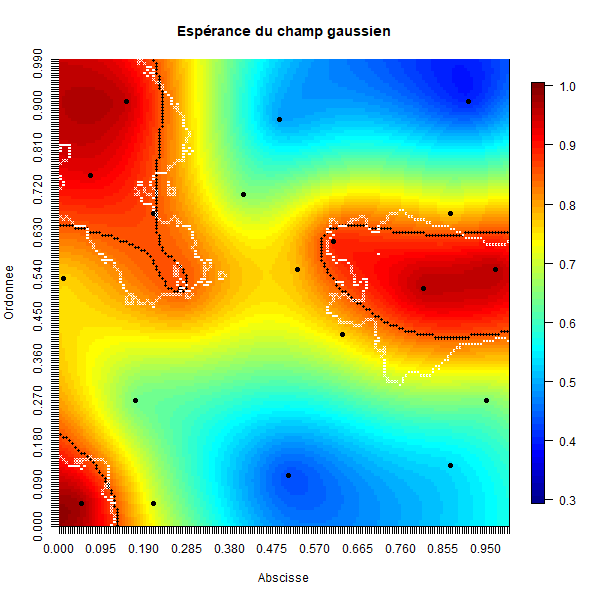}
 				\caption{Map for $\operatorname{IC}^{W}$.}
 			\end{center}
 		\end{subfigure}
 		
 		\caption{comparison of  $\widehat y(x)$  maps   for several $n$-point designs  (black dots) for Example 2. The actual (white)  and estimated (black) level sets are displayed.}
 		\label{fig.complex.10points}	\label{fig.complex.6points}\label{fig.complex.13points}
 	\end{center}
 	
 \end{figure}

 After  $3$   stages, all the   $6$-point designs identify the upper left level set. The 6-point design based on  $\operatorname{MC}^{\mathrm{ls}}$ also identifies a second level set in the right-hand part of the area (see Fig. \ref{fig.complex.6points}).   The designs based on the two integrated criteria need four additional stages to identify a second component (see Fig. \ref{fig.complex.10points}) whereas the design based on $\operatorname{MC}^{\mathrm{W}}$ requires eight additional stages.

 After just  7 stages, the $10$-point design based on $\operatorname{MC}^\mathrm{ls}$ is the only one capable of finding the three components of the level set (see Fig. \ref{fig.complex.10points}). The three components are correctly identified by the designs based on the integrated criteria for the first time with a 14-point design (not displayed in the paper) and by the $\operatorname{MC}^{\mathrm{W}}$ criterion with a 17-point design.

 \begin{figure}
 	\begin{center}
 		\begin{subfigure}[b]{0.32\textwidth}
 			\begin{center}
 				\includegraphics[scale=0.24,trim={30 30 10 40},clip]{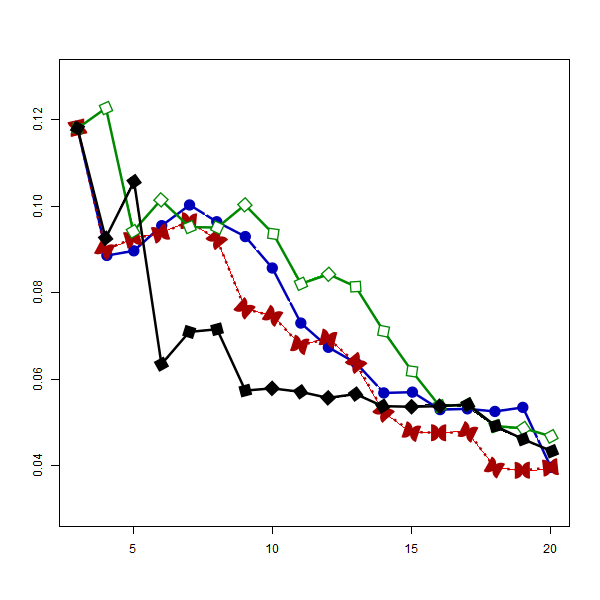}
 				
 				\caption{  $Q_\mathrm{value}$   score}
 			\end{center}
 		\end{subfigure}
 		\begin{subfigure}[b]{0.32\textwidth}
 			\begin{center}
 				\includegraphics[scale=0.24,trim={30 30 10 40},clip]{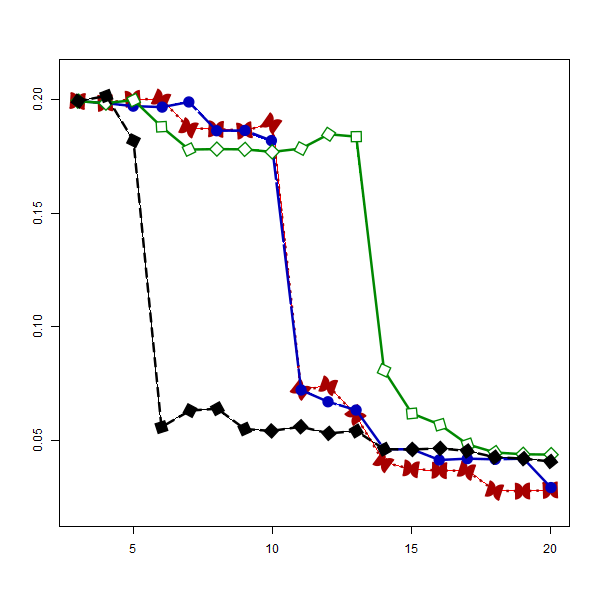}
 				\caption{  $Q_\mathrm{dist}$   score}
 			\end{center}
 		\end{subfigure}
 		\begin{subfigure}[b]{0.32\textwidth}
 			\begin{center}
 				\includegraphics[scale=0.24,trim={30 30 10 40},clip]{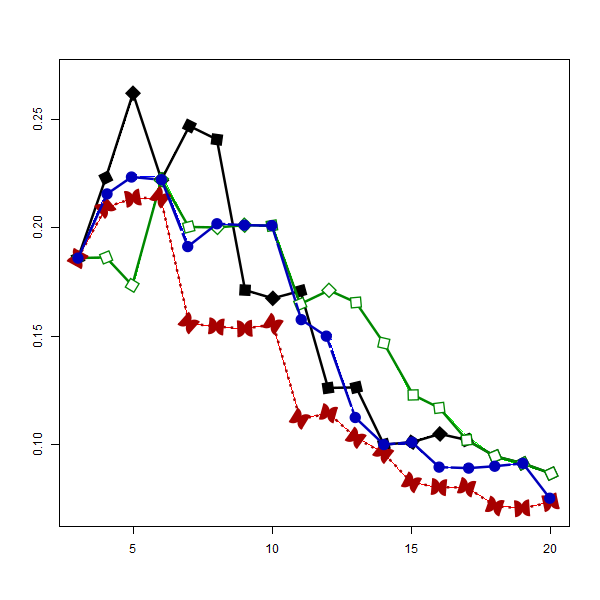}
 				\caption{ $Q_\mathrm{area}$   score}
 			\end{center}
 		\end{subfigure}
 	\end{center}
 	\caption{quality scores against  the number of design points    w.r.t criteria $\operatorname{MC}^{\mathrm{ls}}$ (black), $\operatorname{IC}^{\mathrm{ls}}$ (blue), $\operatorname{IC}^{W}$ (red) and $\operatorname{MC}^{W}$ (green) for Example 2.}
 	\label{fig.score.complex}
 	
 \end{figure}

 The four   $20$-point  designs approximately identify the shape of the three components of the level set. For the three quality scores (Fig. \ref{fig.score.complex}), the two integrated criteria give slightly better results.

 At any stage, the three quality scores consistently show that $\operatorname{MC}^{W}$ provide poor designs. The integrated criteria $\operatorname{IC}^{\mathrm{W}}$ and  $\operatorname{IC}^{\mathrm{ls}}$ provide  the best designs w.r.t. the   quality scores $Q_{\mathrm{dist}}$ and $Q_{\mathrm{area}}$  for more than $14$ points.
 Up to $13$ points, our criterion  $\operatorname{MC}^{\mathrm{ls}}$ provides the best designs  w.r.t. $Q_{\mathrm{value}}$ and $Q_{\mathrm{dist}}$.

 \subsection{Comments}
 The examples presented above show that all four criteria provide highly relevant models for finding smooth level sets. For irregular and disconnected level sets, the models behave very differently, depending on the criterion.

 Sequential designs based on the max-criterion  $\operatorname{MC}^{\mathrm{ls}}$ appear to be more efficient  in detecting disconnected components of the target level set more quickly, especially when they are located close to the domain boundary. Consequently, $\operatorname{MC}^{\mathrm{ls}}$ seems to be the best option in the early stages of sequential design.  Conversely, the maximum criterion $\operatorname{MC}^W$ usually gives poor results.

 Designs based on  the integrated criteria $\operatorname{IC}^{\mathrm{ls}}$ and $\operatorname{IC}^W$ give, in average, similar results.
Overall, the  integrated criteria lead to designs that are  more space filling designs than those based on  maximal criteria: they efficiently control   uncertainty over the whole  area.

 For both examples  several initial designs were tested,  including 3/4/5/6-point space-filling designs and random designs.
 When all the space-filling design  points are located on the same side of the level set, results similar to those presented here were obtained.
 When the points  are located on either sides of the level set, each criterion has resulted in equivalent designs  in terms of quality scores.

 \section{Simulation studies for non-sequential designs}
 \label{section.nonseqstudy}
\begin{figure}
 	\begin{center}
 		\begin{subfigure}[b]{0.48\textwidth}
 			\begin{center}
 				\includegraphics[scale=0.28,trim={30 30 10 60},clip]{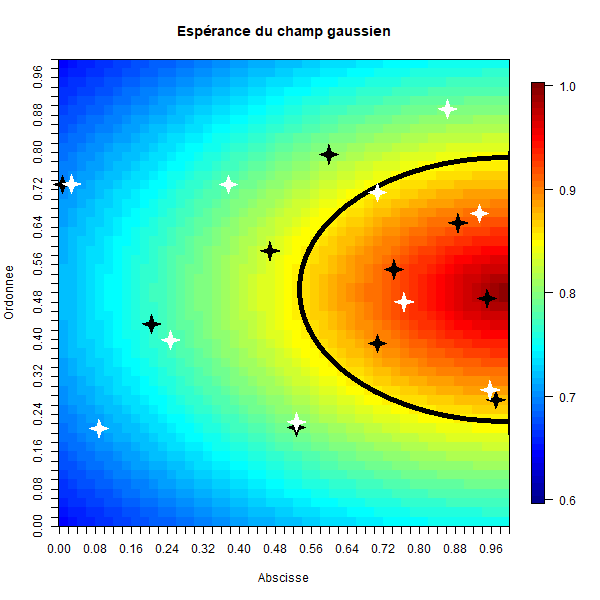}
 				\captionsetup{justification=centering}
 				\caption{10-point designs w.r.t.\\ $\operatorname{MC}^{\mathrm{exc}}$ (black) and $\operatorname{IC}^{\mathrm{exc}}$ (white) }
 				\label{exc.mu.smooth}
 			\end{center}
 		\end{subfigure}
 		\begin{subfigure}[b]{0.48\textwidth}
 			\begin{center}
 				\includegraphics[scale=0.28,trim={30 30 10 60},clip]{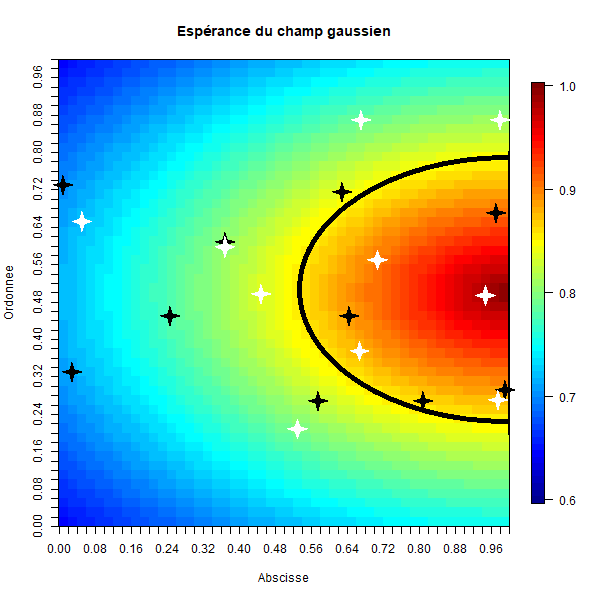}
 				\captionsetup{justification=centering}
 				\caption{10-point designs w.r.t.\\ $\operatorname{MC}^{\mathrm{ls}}$ (black) and $\operatorname{IC}^{\mathrm{ls}}$ (white) }
 				\label{LS.mu.smooth}
 			\end{center}
 		\end{subfigure}
 	\end{center}
 	\vspace*{-3mm}
 	\caption{Map of $\mu(x)$, level set $\{x\,:\,\mu(x)=0.85\}$ (black line) and 10-point designs. }
 	\label{fig.mu.smooth}
 \end{figure}

 In this section, we propose some simulation studies that show how our criterion influence the location of the designs points.
 We consider here  a Gaussian random field $y\sim \mathcal N(\mu,\Sigma)$ on a $50\times50$ grid $E$ over $[0,1]^2$, where $\Sigma$ is  defined, as in Section \ref{section.seqdesignexample} by  \begin{equation}\mathrm{cov}(x_i,x_j)=2^{1-\nu}\sigma^2(d\sqrt{2\nu}/\kappa)^\nu K_\nu(d\sqrt{2\nu}/\kappa)/\Gamma(\nu)
 \end{equation}
 with  $\sigma=0.7$, $\nu=0.7$ and $\kappa=0.2$,  where  $d=d(x_i,x_j)$ is the Euclidean distance between $x_i$ and $ x_j$, $\Gamma$ is the Gamma function and $K$ the modified Bessel function. The mean  function   is given  by the following formula  $\mu(x)= 2 \times \mathrm{exp} \left(-{\{(x_1-1)^2+3(x_2-0.5)^2\}^{1/2}}/{3}\right)$, with $x=(x_1,x_2)$, see Fig. \ref{fig.mu.smooth}. Mean and variance functions can be interpreted as a prior knowledge on  $y(x)$ based, for example, on previous observations or on forecasting models.

 \subsection{Efficient design for area exceeding a given threshold}
 \label{subsection.crit2.result}
 We aim to construct a  $10$-point design that focuses on the area where the expected values exceed a given  threshold $T$. In our example, we choose  $T=0.85$.

 We consider two  criteria,  $\operatorname{MC}^{\mathrm{exc}}_{T}$ and    $\operatorname{IC}^{\mathrm{exc}}_{T}$, obtained from (\ref{eq.Maxcrit}) and (\ref{eq.integratedcrit}) by  choosing the weight function $w^{\mathrm{exc}}_T(x)$.
 For each criterion, we start with the design $d^{\dag}$  obtained by   Algorithm \ref{algo.d0star}. Then, we apply the exchange algorithm with $10\,000$ iterations to obtain the  design $d^*$.
 The resulting designs are displayed in Fig. \ref{exc.mu.smooth}.

\begin{figure}
 	\begin{center}
 		\begin{subfigure}[b]{0.48\textwidth}
 			\begin{center}
 				\includegraphics[scale=0.20,trim={25 30 10 50},clip]{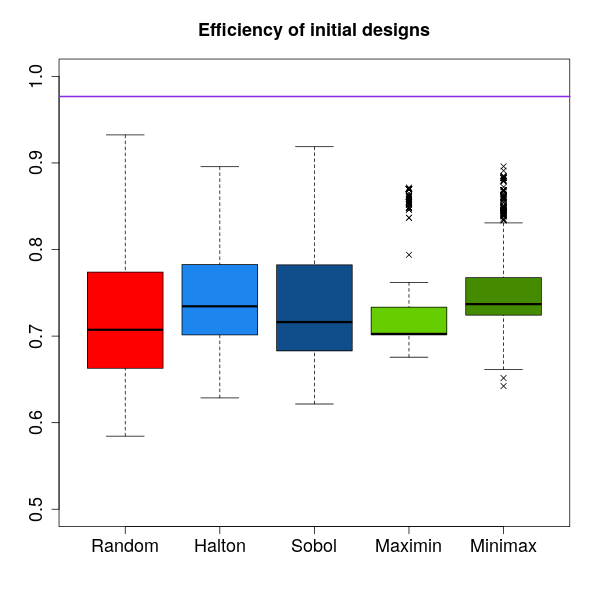}
 				\caption{for $\operatorname{MC}^{\mathrm{exc}}_{T}$.}
 				\label{districritinit1}
 			\end{center}
 		\end{subfigure}
 		\begin{subfigure}[b]{0.48\textwidth}
 			\begin{center}
 				\includegraphics[scale=0.20,trim={25 30 10 50},clip]{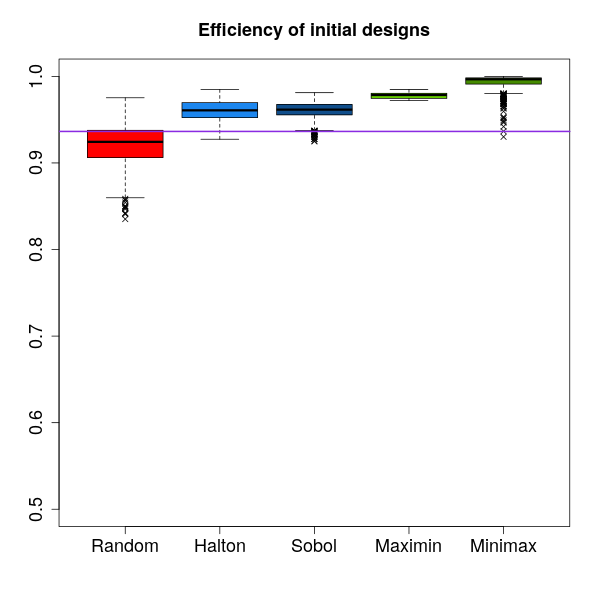}
 				\caption{for $\operatorname{IC}^{\mathrm{exc}}_{T}$.}
 				\label{districritinit2}
 			\end{center}
 		\end{subfigure}
 	\end{center}
 	\vspace*{-3mm}
 	\caption{Efficiency factors of usual designs w.r.t. $\operatorname{MC}^{\mathrm{exc}}_{T}$ and $\operatorname{IC}^{\mathrm{exc}}_{T}$. The purple line corresponds to the efficiency of $d^{\dag}$.}
 	\label{districritinit}
 \end{figure}

 In Fig. \ref{districritinit1}, we compare the efficiency factors   (\ref{eq.efficiency})  of the designs $d^{\dag}$, $d^*$ and usual space filling  designs. Space filling designs  are obtained    from the R packages \texttt{randtoolbox}, \texttt{minimaxdesign} and \texttt{maximin}. As these packages use optimization algorithm  which include  a degree of randomness, we   ran times each package several. For each type of space filling designs, we display the    box plots of the efficiency factors.

 We can see that the starting design $d^{\dag}$ is highly efficient w.r.t. both   $\operatorname{MC}^{\mathrm{exc}}_{T}$ and $\operatorname{IC}^{\mathrm{exc}}_{T}$.
 We can also observe on our example that  space filling designs are less efficient  against  $\operatorname{MC}^{\mathrm{exc}}_{T}$ than  w.r.t.  $\operatorname{IC}^{\mathrm{exc}}_{T}$. As it will be seen in Section \ref{section.seqdesignexample}, this illustrates the fact that integrated criteria lead to  more space filling designs than max-criteria.

 \subsection{Efficient design to detect level sets}
 \label{subsection.crit3.result}

 We aim to estimate the level set $y(x)=T$ with $T=0.85$.
 We  use the same meta-model  as in Section \ref{subsection.crit2.result}  with the same parameters. So, we seek the optimal $10$-point design w.r.t. the $\operatorname{MC}^{\mathrm{ls}}_{T}$ and $\operatorname{IC}^{\mathrm{ls}}_{T}$ criterion, where
 $$\operatorname{IC}^{\mathrm{ls}}=\sum_{x\in E}w^{\mathrm{ls}}_T(x)\times \mathrm{Var}\left(y(x)|d\right)$$
 or
 $$\operatorname{MC}^{\mathrm{ls}}=\max_{x\in E}\{w^{\mathrm{ls}}_T(x)\times \mathrm{Var}\left(y(x)|d\right)\}.
 $$

 \begin{figure}
 	\begin{center}
 		\begin{subfigure}[b]{0.48\textwidth}
 			\begin{center}
 				\includegraphics[scale=0.20,trim={25 30 10 50},clip]{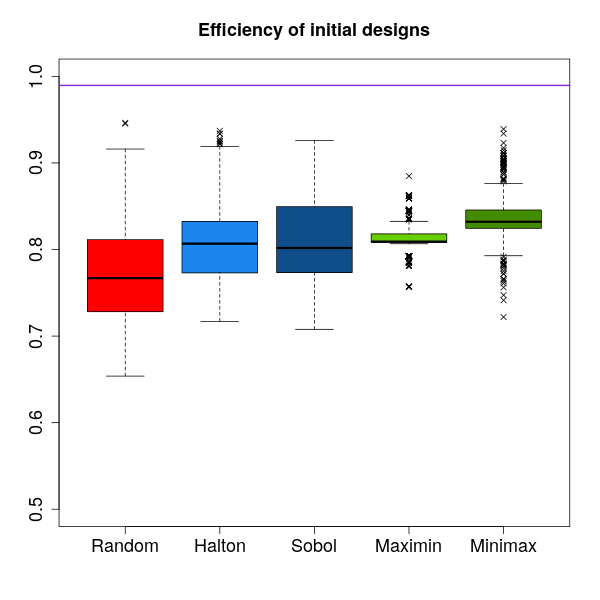}
 				\caption{for $\operatorname{MC}^{\mathrm{ls}}_{T}$.}
 				\label{districritinit3}
 			\end{center}
 		\end{subfigure}
 		\begin{subfigure}[b]{0.48\textwidth}
 			\begin{center}
 				\includegraphics[scale=0.20,trim={25 30 10 50},clip]{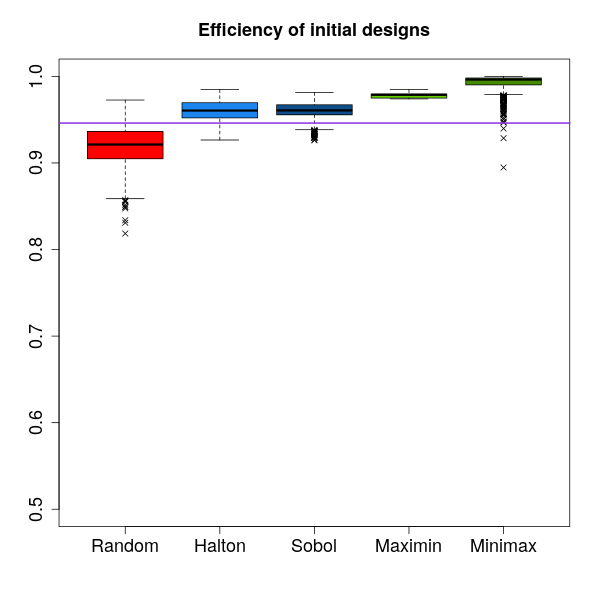}
 				\caption{for $\operatorname{IC}^{\mathrm{ls}}_{T}$.}
 				\label{districritinit4}
 			\end{center}
 		\end{subfigure}
 	\end{center}
 	\vspace*{-3mm}
 	\caption{Efficiency factors of usual designs w.r.t. $\operatorname{MC}^{\mathrm{LS}}_{T}$ and $\operatorname{IC}^{\mathrm{LS}}_{T}$. The purple line corresponds to the efficiency of $d^{\dag}$.}
 	\label{districritinitls}
 \end{figure}

 Clearly, a sequential design is far  more appropriate to this situation, since observations are needed to estimate the level set (see Section \ref{section.seqdesignexample}). However, a non-sequential design can be seen as   a stage  of a group-sequential design   based on the   meta-model updated with the previous observation.

 ~

 The resulting designs are displayed in Fig. \ref{LS.mu.smooth}. In Fig. \ref{districritinitls}, we compare their efficiency factors  with usual space filling designs. Similarly to  Section \ref{subsection.crit2.result}, designs  $d^{\dag}$ obtained by Algorithm  	\ref{algo.d0star}
 are highly efficient  with an efficiency factor   $0.993$ for $\operatorname{MC}^{\mathrm{ls}}$ and  $0.94$ for $\operatorname{IC}^{\mathrm{ls}}$. In this example, space-filling designs are highly efficient for $\operatorname{IC}^{\mathrm{ls}}$, but not for the $\operatorname{MC}^{\mathrm{ls}}$ criterion.

Further experiments were carried out by alternately considering $3$-point designs and $20$-point designs. As with $10$-point designs, $d^{\dag}$ is more relevant than space-filling designs (in particular for the $\operatorname{MC}^{\mathrm{ls}}$ criterion) and is slightly improved by the exchange algorithm. However, as the number of design points increases, the differences of efficiencies  between the designs  are reduced.

 \section{Conclusion}
 \label{section.conclusion}

 We  proposed  new criteria  to build optimal designs that aim to accurately estimate the response over  a  given  target area when the response is modelled by a Gaussian field. From a Bayesian point of view, these criteria are based on the faith the a point belong or not to the  are of interest. From a  frequentist point of view, they are based on the test of whether the point belongs to the target area or not.

 When the aim of the experiment is to estimate  a level set, we have  proposed three quality scores to evaluate and compare  the performance of the designs.
 In the case of sequential designs,   the max-criterion $\operatorname{MC}^{\mathrm{ls}}_{T}$ appears to be more effective than the other criteria in exploring areas of  high uncertainty and therefore to detect disconnected areas. Integrated criteria are more space filling and attempt to control overall uncertainty, with larger weights on the already explored   areas.
 This suggests  a  hybrid strategy for  investigating a target area using sequential designs: first, use the  max-criterion  $\operatorname{MC}^{\mathrm{ls}}$ to quickly identify areas of interests. Then, after several stages,   use  the  integrated criterion  $\operatorname{IC}^{\mathrm{ls}}$ or  to     accurately  control  the overall uncertainty on that area. The evaluation of the performance of this approach is left for future research.

For non-sequential designs,   optimal designs are highly dependent on prior information, so there is no natural way of comparing   designs with each other. For this situation, we proposed a  non-optimal but  efficient design,  $d^\dag$,  based on a computationally-cheap algorithm.

\subsection*{Acknowledgment}

This research was financed by the French government IDEX-ISITE initiative 16-IDEX-0001 (CAP 20-25).

\bibliographystyle{elsarticle-num-names}  
\bibliography{biblioplandexp}

\end{document}